\documentclass[
reprint,amsmath,amssymb,aps,pra,floatfix,longbibliography]{revtex4-1}
\usepackage{graphicx}	% Include figure files
\usepackage{dcolumn}	% Align table columns on decimal point
\usepackage{bm}            % bold math
\usepackage{color}

\begin{document}
%\preprint{APS/123-QED}
\title{
Matrix isolated 
barium 
monofluoride:
Assembling a 
sample of 
BaF
molecules for
a measurement 
of the 
electron
electric dipole
moment
}
\author{Z. Corriveau}
\author{R.L. Lambo}
\author{D. Heinrich}
\author{J. Perez Garcia}
\author{N.T. McCall}
\author{H.-M. Yau}
\author{T. Chauhan}
\author{G.K. Koyanagi}
\author{A. Marsman}
% \author{R. Fournier}
\author{M.C. George}
\author{C.H. Storry}
\author{M. Horbatsch}
\author{E.A. Hessels} 
\email{hessels@yorku.ca}
\affiliation{Department of Physics and Astronomy, York University, Toronto,
Ontario M3J 1P3, Canada}

\date{\today} % It is always \today, today, but any date may be explicitly specified

\begin{abstract}

A cryogenic neon solid
doped with 
barium 
monofluoride 
(BaF)
is created
on a cryogenic substrate
using 
a stream of 
Ne gas 
and 
a 
high-intensity
beam of
BaF 
molecules produced in
a 
cryogenic
helium-buffer-gas
laser-ablation source.
The apparatus  
is designed for 
eventual use in 
a measurement of 
the electron
electric dipole moment
(eEDM). 
Laser-induced fluorescence
is observed from
transitions up to the 
$B\,^2\Sigma_{1/2}$
state.
The number of 
BaF
molecules 
found to 
be present in the solid
and addressable with this 
laser transition
is
approximately
10$^{10}$
per
mm$^3$,
which is 
of the same order as the total
number of 
BaF 
molecules  
that impact 
% each 
% mm$^2$
% of 
the substrate
during the 
hour of growth time for the solid.
As a result, 
an eventual 
eEDM 
measurement
could have 
continual
access 
to 
an 
accumulation of 
an 
hour's 
worth of molecules
(all of which 
are contained within 
a
1-mm$^3$
volume
and
are thermalized
into the ground state),
compared to beam experiments 
which study 
the molecules from a single
ablation during the 
millisecond-timescale
of their passage
through a 
much
larger-scale
measurement apparatus.
The number 
of
BaF
molecules 
observed in the 
doped solid 
matches the target value for 
our planned 
eEDM
measurement.
%\begin{description}
% \item[Usage]
% Secondary publications and information retrieval purposes.
% \item[PACS numbers]
% \verb+\pacs{32.70.Jz,32.80.-t}+
%\end{description}

\end{abstract}

\pacs{Valid PACS appear here}% PACS, the Physics and Astronomy Classification Scheme.
\maketitle

\section{Introduction}
The barium 
monofluoride
(BaF)
molecule
has been extensively studied
both experimentally 
\cite{
% mitsherlich1864,
% george1913,
% datta1921spectra,
walters1928alkaline,
Johnson1929band,
nevin1931spectrum,
jenkins1932emission,
fowler1941new,
ehlert1964mass,
hildenbrand1968mass,
knight1971hyperfine,
kushawaha1972green,
kushawaha1973c2pi,
dagdigian1974radiative,
ryzlewicz1980formation,
Ip1981OpticalOptical,
ryzlewicz1982rotational,
ernst1986hyperfine,
effantin1987laser,
barrow1988metastable,
effantin1990studiesI,
bernard1990studiesII,
bernard1992laser,
berg1993lifetime,
jakubek1994core,
jakubek1994ionization,
guo1995high,
jakubek1996core,
jakubek1997rydberg,
berg1998lifetime,
jakubek2001core,
Steimle2011MolecularBeam,
cahn2014zeeman,
zhou2015direct,
Bu2017Cold,
cournol2018rovibrational,
aggarwal2019lifetime,
aggarwal2021supersonic,
bu2022saturated,
courageux2022efficient,
mooij2024influence,
touwen2024manipulating}
and 
theoretically
\cite{
torring1989energies,
allouche1993ligand,
arif1997rydberg,
kobus2002comparison,
tohme2015theoretical,
prasannaa2016permanent,
bala2019ab,
haase2020hyperfine,
skripnikov2021role,
denis2022benchmarking,
kyuberis2024theoretical},
with recent interest 
centred on the potential
of 
BaF
molecules being used for
laser manipulation,
cooling,
and 
trapping
\cite{
Chen2016Structure,
chen2016erratum,
Kang2016Suitability,
xu2017baf,
chen2017radiative,
hao2019high,
liang2019improvements,
albrecht2020buffer,
yang2020ab,
kogel2021laser,
zhang2022doppler,
marsman2023large,
marsman2023deflection,
rockenhauser2023absorption,
rockenhauser2024laser,
zeng2024three,
kogel2024isotopologue}
and for 
a measurement of 
the violation of 
parity- 
and 
time-reversal symmetries
\cite{
kozlov1985semiempirical,
kozlov1997enhancement,
meyer2006candidate,
nayak2006ab,
nayak2008calculation,
nayak2009relativistic,
flambaum2014time,
fukuda2016local,
hao2018nuclear,
altuntacs2018demonstration,
abe2018application,
vutha2018oriented,
vutha2018orientation,
aggarwal2018measuring,
prasannaa2019role,
denis2020enhanced,
talukdar2020relativistic,
haase2021systematic,
prosnyak2024axion,
boeschoten2024spin}.
The 
EDM$^3$ 
collaboration 
is endeavouring 
\cite{
vutha2018orientation,
vutha2018oriented}
to use
BaF
molecules embedded 
in an 
inert-gas 
matrix
to measure the 
electron electric 
dipole moment
(eEDM).
Confining the 
BaF 
molecules within
a cryogenic matrix has the  
advantage of producing 
a much larger sample of 
molecules
than conventional 
eEDM
measurements
\cite{hudson2011improved,
baron2013order,
Cairncross2017,
acme2018improved,
roussy2023new},
which use 
molecular beams
or ion traps.
Matrix isolation is expected
to also reduce
systematic uncertainties
for an 
eEDM 
measurement
because of the small
volume and cold
temperature of the 
sample of molecules
and 
because the matrix
orients the molecules
(without the need for 
an external electric field)
and keeps the molecules 
stationary
\cite{
lambo2023calculationAr,
lambo2023calculationNe}.

Two previous experiments 
\cite{
knight1971hyperfine,
li2023optical}
studied  
matrix-isolated 
BaF 
molecules,
both in a neon matrix,
and detailed theoretical
predictions for the 
matrix-isolated 
BaF
molecules
have been made
\cite{
koyanagi2023accurate,
lambo2023calculationAr,
lambo2023calculationNe}
by our 
EDM$^3$
collaboration.
In this work, 
we describe the apparatus 
and methods for producing 
a 
BaF-doped
Ne solid 
from a beam of 
BaF
molecules.
The
apparatus is designed to be suitable for 
an eventual measurement of the 
eEDM.
We
study the  
laser-induced 
fluorescence
that results from excitation 
to the 
$B\,^2\Sigma_{1/2}$
state
and
establish lower limits on the concentration
of 
$^{138}$BaF 
molecules in the 
Ne
solid.
We find that the 
solid 
efficiently
collects and isolates
BaF
molecules from the beam,
and that our solid 
has approximately
$10^{10}$
BaF 
molecules per
mm$^3$,
which is 
the target value for 
our planned measurement 
of the 
eEDM.

% find that the 
% laser excitation
% and fluorescence spectra
% of 
% BaF 
% in 
% Ne 
% for the first five excited states
% (the
% A$^\prime\,^2\Delta_{3/2}$,
% A$^\prime\,^2\Delta_{5/2}$,
% A$\,^2\Pi_{1/2}$,
% A$\,^2\Pi_{3/2}$,
% and
% B$\,^2\Sigma_{1/2}$
% states,
% which form
% \cite{bernard1990studiesII}
% the 
% 5d
% complex of states)
% becomes 
% quite simple after annealing,
% and that a study of the 
% decay back down to the ground
% electronic state 
% can be used to deduce properties
% of the 
% BaF:Ne
% system.
% Most importantly for the 
% EDM$^3$
% collaboration, 
% there are
% strong
% evidence that 
% the vast majority the
% BaF
% molecules
% are
% in a single substitution site.

\section{Experimental Procedure}

\begin{figure}
\includegraphics
[width=1.0\linewidth]
{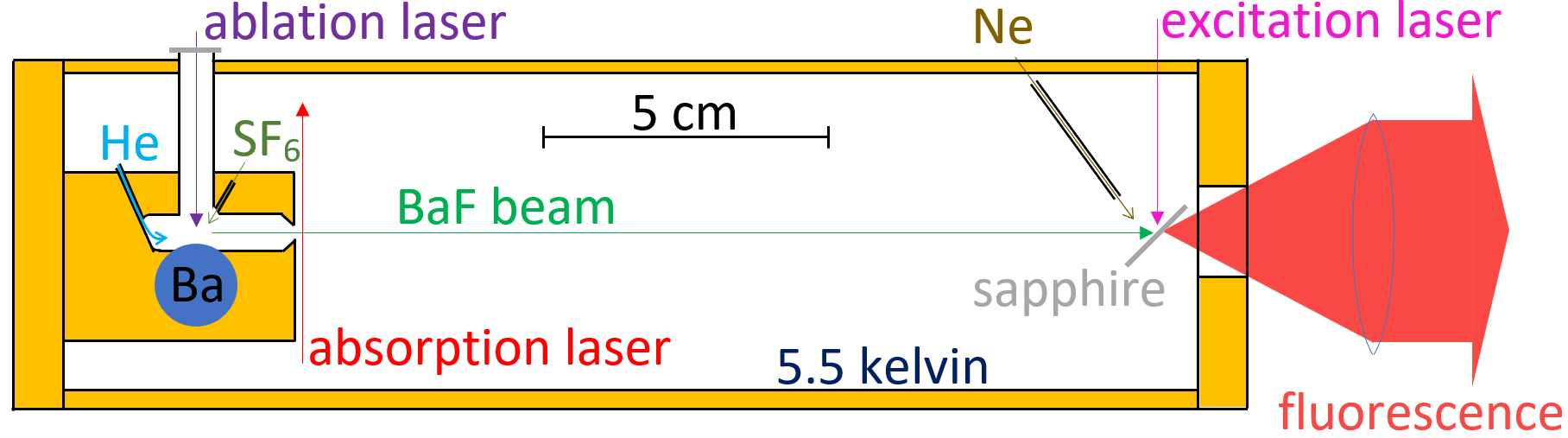}
\caption{
\label{fig:setup}
A cross-sectional view of the 
cryogenic experimental 
apparatus.
BaF 
molecules are created
by laser ablation of 
barium metal in a 
helium-buffer-gas cell,
with a small stream of 
SF$_6$ 
present.
A sapphire substrate 
located
20~cm 
from the output of the 
buffer-gas
cell 
has a small flow of 
neon gas directed at it
in addition to the beam 
of 
BaF 
molecules.
After producing and annealing 
the doped neon solid, 
laser induced fluorescence 
is observed.
}
\end{figure}

BaF
molecules are created 
within a 
helium-buffer-gas 
cell via laser ablation of 
barium metal in the 
presence of a 
small stream 
of
SF$_6$
gas,
using 
a design based on the one used by the 
group at 
Imperial College
\cite{truppe2018buffer}.
Fig.~\ref{fig:setup}
shows 
(to scale)
a 
cross-sectional 
view of the 
10-mm-diameter
buffer-gas cell,
which is machined 
from a solid block
of
oxygen-free 
high thermal conductivity 
(OFHC) 
copper.
This copper
block is 
thermally anchored
to a surrounding 
5-mm-thick
OFHC 
copper
box
(see 
Fig.~\ref{fig:setup}),
which is
cooled to 
5.5~K
using a 
Sumitomo
RDK-408D2   
cryogenic 
refrigerator.
This box has removable
panels and its
inside is 
coated with 
activated coconut charcoal
which efficiently 
cryopumps
excess helium gas.
This
5-K
copper box is surrounded by a  
slightly larger
aluminium box
(not shown in 
Fig.~\ref{fig:setup}),
with 
sapphire windows for 
optical access,
which
is maintained at a temperature
of 
60~K 
to shield the inner box from 
room-temperature
blackbody 
radiation.
The thermal connection
from the 
two-stage 
Sumitomo
refrigerator
to both 
boxes 
is via several very flexible
57-mm-long
braided 
OFHC 
copper 
straps
(Technology Applications, 
Inc. 
model 
P50-502)
which isolate the apparatus
from the vibrations inherent
in the refrigerator.

A 
20-mm-diameter
barium rod is situated 
at the bottom of this 
buffer-gas 
cell.
The rod can be translated 
(into the page in 
Fig.~\ref{fig:setup})
and can be rotated 
to allow a fresh
surface of barium
be exposed to the 
ablation laser.
The barium metal is 
ablated using
6-ns-long,
1064-nm
laser pulses
from 
a 
Litron Lasers
model 
Nano-L 90-100
Nd:YAG
laser
which is capable of 
producing up to 
100
pulses per second and 
up to 
90~mJ
per pulse.
The ablation laser beam 
travels through a 
10-mm-diameter,
53-mm-long
nozzle which limits the contamination 
of the laser entrance window.

A helium flow of
10~standard
cubic centimeters per minute
(sccm)
flows through this cell 
during laser ablation.
A small constant flow
(0.06~sccm)
of 
SF$_6$
gas
is used during ablation,
allowing the ablated 
barium atoms to 
react and form
the 
BaF
radical.
The 
BaF
molecules are quickly
cooled by the buffer
gas and are entrained 
in the helium flow.
The resulting beam of 
BaF 
molecules
exits the 
3.5-mm-diameter
nozzle with a 
forward speed 
similar to the helium
atoms
(ranging from 
150
to
300~m/s,
depending on the 
flow rate of helium
and the temperature of
the 
buffer gas).
The 
buffer-gas
cell 
heats up to 
% is maintained 
% at 
a temperature of
6.8~K
during the 
laser-ablation
process,
but
contamination of the 
inner surface of the 
cell can form an insulating 
layer, 
resulting in a helium 
temperature that is
warmer than the temperature
of the cell.

The 
burst of  
BaF 
molecules produced by
each ablation pulse 
is monitored via
laser absorption, 
using a 
diode laser beam
(shown in 
Fig.~\ref{fig:setup},
directed perpendicular
to the 
BaF
beam direction,
1.5~mm
away from the 
output nozzle 
of the 
buffer-gas
cell,
$1/e^2$
diameter of 
1.75~mm,
power of 
10~$\mu$W,
and
central intensity
of 
0.21~mW/cm$^2$)
tuned to 
859.839 nm,
which is in resonance 
with the transition from 
the 
$^{138}$BaF
lowest 
vibrational
($v$$=$$0$)
and
second-lowest
rotational 
($N$$=$$1$)
state
of the 
$X\,^2\Sigma^+_{1/2}$ 
ground electronic state
to the 
positive-parity
$A\,^2\Pi_{1/2}$
($v$$=$$0$,
$j$$=$$1/2$)
state.
This transition  
is a cycling transition
(in that it does not 
lose population to other
rotational states)
and is used 
for laser cooling and 
manipulation of 
BaF
molecules
\cite{
Chen2016Structure,
chen2016erratum,
Kang2016Suitability,
xu2017baf,
chen2017radiative,
hao2019high,
liang2019improvements,
albrecht2020buffer,
yang2020ab,
kogel2021laser,
zhang2022doppler,
marsman2023large,
marsman2023deflection,
rockenhauser2023absorption}.
In the present case,
however,
our 
low-intensity,
single-frequency
laser interacts with 
only one of the four
$v=0, 
N=1$
hyperfine levels
and
the population is quickly
pumped to the other hyperfine
levels,
limiting the number of absorbed
photons for each molecule 
to of order unity.
The 
Doppler width
due to a
transverse velocity 
in our beam 
(FWHM
of approximately
50 MHz,
compared to the 
natural width of 
2.8~MHz
for  
this transition)
implies 
that only a small
fraction 
(approximately 
5\%)
can interact with the 
single-frequency laser.
At the temperature of our
buffer cell, 
the vast majority 
of the 
BaF 
molecules will be 
in the 
$v$$=$$0$ 
vibrational
state,
with 
the 
fraction of the population
in the 
$N$$=$$1$
rotational
state 
being determined 
by the thermal
distribution 
at the 
temperature of the 
buffer gas
(approximately 
20\% 
at our temperatures).

Density-matrix 
calculations 
(similar to those in 
Refs.
\cite{marsman2023deflection,
marsman2023large})
are employed to estimate
the 
average
number of photons
absorbed per 
BaF
molecule.
These simulations
include 
a full
Monte-Carlo
simulation 
for the range
of 
velocities 
(based on 
an 
8.5-kelvin 
buffer-gas
temperature,
and 
a forward 
speed 
of 
250~m/s, 
which is inferred 
from the time delay
between observation
points along the 
20-cm
path of the 
BaF
molecules)
and 
initial positions
of the 
BaF
molecules as they
exit the 
buffer-gas 
cell.
This modelling
uses full
numerical integration 
of the 
density-matrix
equations for the 
transit time through
the laser beam.
The simulations
give an overall estimate
of 
0.049
absorbed photons 
per
BaF 
molecule
exiting the 
buffer-gas 
source
in the
$v$$=$$0, N$$=$$1$
state.

This modeled
absorption can 
be compared to the
observed
absorption,
which has 
an amplitude of 
2\%
(0.2~microwatt)
and a duration of 
50~$\mu$s
(the emptying time 
of the cell),
for a total of 
approximately 
40~million
photons
absorbed per ablation
pulse.
The comparison implies 
that there are 
900~million 
molecules in the 
$v$$=$$0, 
N$$=$$1$
state,
and a total of 
4~billion
$^{138}$BaF 
molecules 
produced per
ablation pulse.
This estimate leads to 
30~billion
$v$$=$$0, 
N$$=$$1$
molecules
per 
steradian
per
ablation pulse
(within a factor 
of
two
of the number
found in 
Ref.~\cite{truppe2018buffer},
where the number of 
CaF 
molecules in a similar
source was estimated
by observing fluorescence
while exciting all four
hyperfine 
states of the cycling transition
using 
radio-frequency 
sidebands).
Including all 
rotational states,
the total production of 
BaF
molecules
is estimated to be
140~billion
total 
per 
steradian
per
ablation pulse.

A total of 
approximately
$20\,000$~ablation pulses
over a time period of
about 
1~hour
are used.
A 
1-mm-thick
sapphire substrate
is located 
20~cm
downstream of this source.
This 
20-cm 
spacing will allow for 
separation of 
BaF 
molecules from other 
ablation products.
Modelling has shown that this 
separation is possible with either
laser deflection
\cite{
marsman2023large,
marsman2023deflection}
or
electrostatic deflection
\cite{yau2024specular}.

The 
BaF
beam 
together with a 
stream of 
20~sccm of Ne
is used to grow a doped
solid that 
has a thickness that 
is of the order of 
1~mm.
During this growth,
the sapphire is 
kept at a temperature
of
6.8~K.
Based on our estimated
BaF
flux,
a total of 
$5\times10^{10}$~BaF
molecules
per 
mm$^2$ 
are directed at the 
solid 
during the growth.
If all of these molecules
are embedded in the solid,
this would correspond to a
BaF:Ne 
ratio
of 
approximately 
1~part per billion.
% ,
% which is the target 
% value for a measurement 
% of the eEDM.

After growth,
the solid is cooled to 
5.8~K.
Several annealing cycles are 
used, 
with the temperature raised
in one minute up
to 
8.4~K,
left at this temperature
for 
5~minutes,
and then 
recooled
in 
one minute
back down to 
5.8~K.

The annealed,
doped solid is
illuminated with a 
tightly focused laser
beam 
% (with tunable wavelength from
% 700
% to
% 1000~nm,
(1/$e^2$
laser diameter
of
40~microns,
laser power of up
to 
0.5~W,
and 
beam-center 
intensity of
up to
20~kW/cm$^2$)
to induce fluorescence.
Approximately 
2.5\%
of the fluorescence
(4\% of  
$4\pi$ 
steradians, 
with 
35\%
loss)
is 
collimated 
by a 
5-cm-diameter lens 
located 
6~cm 
beyond the substrate
(see 
Fig.~\ref{fig:setup}).
The collected light 
passes
% passing through the 
% sapphire substrate 
% on which the solid is
% grown, 
% through a 
% sapphire window 
% on the 
% 60-K 
% thermal box,
out of a  
vacuum window
and,
after 
wavelength
filtering,
is 
refocused.
The intensity of fluorescence
is found to increase 
approximately 
linearly
with time during the growth
of our solid, 
indicating that the solid
remains sufficiently transparent
for transmission of the laser and 
fluorescent light.

The 
refocused
fluorescence is observed by  
one of two spectrometers
(Ocean Optics QE Pro, 
350
to 
1125~nm;
Optosky ATP8200,
900
to
2500~nm),
by a 
3-mm-diameter
liquid-nitrogen-cooled
InGaAs 
photodiode,
or
by a 
single-photon
counter
(Excitas 
SPCM-AQRH-16-FC, 
used in conjunction with
a 
Cronologic
model 
4-2G
time tagger).

\section{Laser-induced
fluorescence}

\begin{figure}
\includegraphics
[width=0.4\linewidth]
{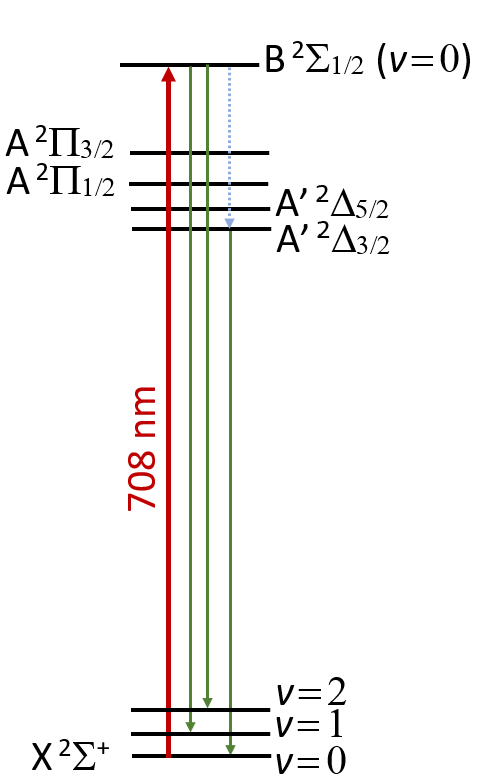}
 \caption{
 \label{fig:energyLevels}
 The 
 relevant electronic and 
 vibrational
 energy levels of 
 BaF. 
 The 
 $X\,^2\Sigma_{1/2} (v$$=$$0)$,
 state
 is 
 laser-excited 
 up to
 the 
 $B\,^2\Sigma_{1/2} (v$$=$$0)$
 state.
 Fluorescence from 
 radiative 
 decay
 down to the 
 $X\,^2\Sigma_{1/2} (v$$=$$1)$
 and
 $X\,^2\Sigma_{1/2} (v$$=$$2)$
 states is observed 
 (while it is not possible to 
 observe fluorescence from 
 decay to 
 $v$$=$$0$,
 as it is too close in
 wavelength to the excitation laser).
 Additionally,
 fluorescence from the 
 $A^\prime\,^2\Delta_{3/2} (v$$=$$0)$
 state is observed. 
 The latter fluorescence 
 results from a 
 nonradiative
 decay from the 
 $B\,^2\Sigma_{1/2} (v$$=$$0)$
 state to the
 $A^\prime\,^2\Delta_{3/2} (v$$=$$0)$
 state. 
}
\end{figure}

\begin{figure}[b!]
\vspace{5mm}
% \begin{subfigure}
    \includegraphics[width=3in]
    {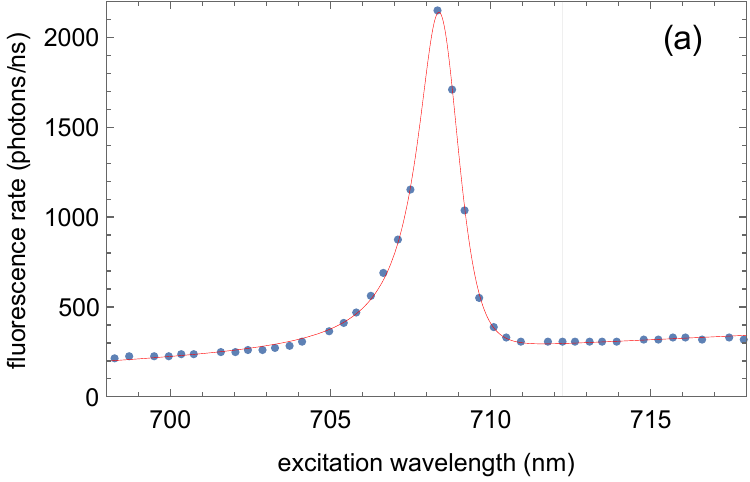}
% \end{subfigure}

\vspace{5mm}
% \begin{subfigure}
    \includegraphics[width=3in]
    {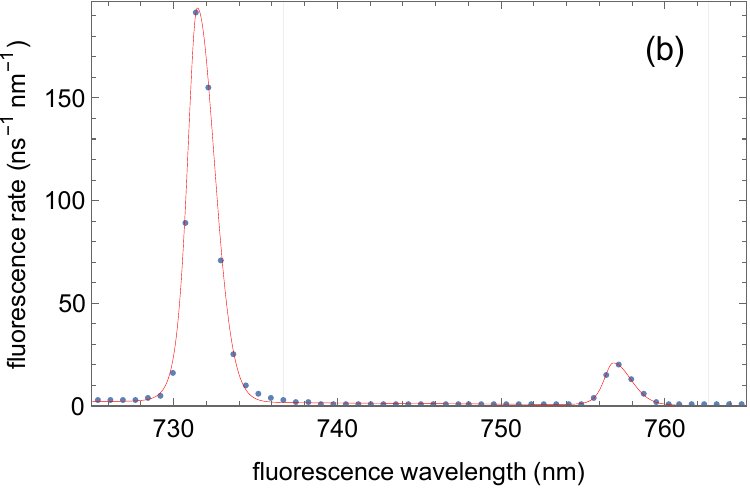}
% \end{subfigure}

\vspace{5mm}
% \begin{subfigure}
    \includegraphics[width=2.8in]
    {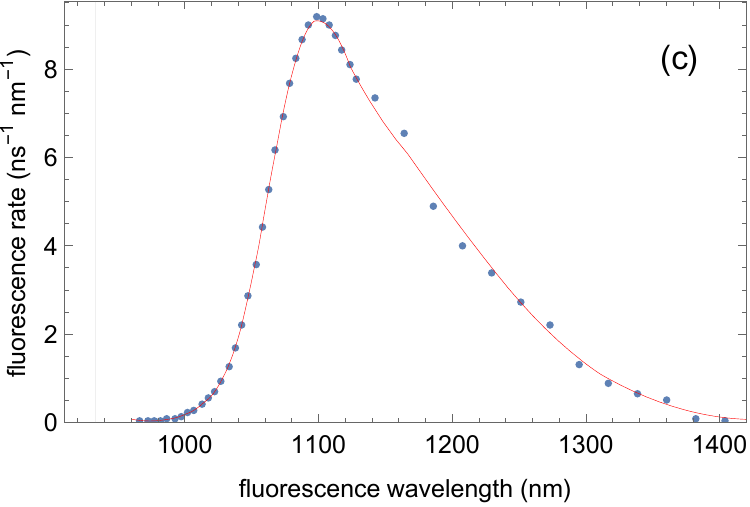}
% \end{subfigure}
\caption{\label{fig:spectra}
The excitation spectrum
(a) 
versus
wavelength for excitation
to the 
$B\,^2\Sigma_{1/2} (v=0)$
and the resulting 
fluorescence spectra
(for excitation at 
708.4~nm) 
for direct decay back to the 
ground electronic state
(b)
and indirect decay 
through the 
$A^\prime\,^2\Delta_{3/2}$
state 
(c).
In 
(b),
decay to both the 
$v=1$
and
$v=2$
vibrational
levels is observed, 
but 
the decay to the
$v=0$
state
is not seen as it 
is too close to the 
excitation wavelength
and is obscured by 
scattered laser light.
The vertical lines 
represent the positions 
of the resonances 
for a free
BaF
molecule 
and the solid curves 
are included to guide the eye.
}
\end{figure}

When the excitation laser is 
tuned near 
708~nm
(see 
Fig.~\ref{fig:energyLevels})
it causes fluorescence  
in several wavelength ranges 
between 
700
to 
1400~nm.
The 
vast majority 
($>$99\%)
of this fluorescence
is found in
two fluorescence peaks: 
one spanning 
730
to
736~nm,
due to  
decay from the 
$B\,^2\Sigma_{1/2}$
and
the other
much broader 
peak 
spanning
980
to
1400~nm,
due to radiative
decay from the 
$A^\prime\,^2\Delta_{3/2}$
state.
The fluorescence versus excitation
wavelength for these two
fluorescence peaks is shown in 
Fig.~\ref{fig:spectra}(a).
The tail on the blue side of the
peak results from laser excitation
to phonon states above the 
$B\,^2\Sigma_{1/2}$
state.
The laser excitation in the 
matrix is shifted
by 
4~nm
to the blue
of the resonance
position for 
a free
$^{138}$BaF 
molecule
(the vertical
line in 
Fig.~\ref{fig:spectra}(a)).

The spectra for the fluorescence
for 
radiative 
decay 
from the 
$B\,^2\Sigma_{1/2}$
and
$A^\prime\,^2\Delta_{3/2}$
states
are shown in 
Fig.~\ref{fig:spectra}(b)
and
(c),
respectively.
The tails on the red 
side of these peaks result
from decay to 
phonon states above the 
$X\,^2\Sigma^+_{1/2}$
state.
The 
$B\,^2\Sigma_{1/2}$-to-$X\,^2\Sigma^+_{1/2}$
fluorescence
(for decay to both the 
$v=1$
and
$v=2$
states)
is shifted 
to the blue
(relative to the free molecule)
by 
5~nm,
whereas the
$A^\prime\,^2\Delta_{3/2}$-to-$X\,^2\Sigma^+_{1/2}$
fluorescence
is shifted to the red by 
150~nm
and is
strongly broadened.

The equilibrium 
positions of the 
BaF 
molecule 
and its surrounding 
Ne
atoms
are
different for 
the 
$B\,^2\Sigma_{1/2}$
state
than for
the 
ground electronic state
(for which the positions 
were examined
theoretically in 
Ref.~\cite{lambo2023calculationNe}).
Thus, 
after laser excitation,
the molecule and surrounding 
atoms quickly make small changes to their
positions.
This movement transfers energy 
to the solid.
On a 
time scale that is much
faster than 
radiative
decay down to the ground state,
the molecule 
(and surrounding atoms)
either settle 
to their equilibrium positions
for the 
$B\,^2\Sigma_{1/2}$
state 
(leading to the fluorescence
of 
Fig.~\ref{fig:spectra}(b))
or, 
we hypothesize,
in the process of settling, 
they experience a level 
crossing that 
leaves them in   
the 
lowest-lying
excited
electronic
state 
of the 
molecule
(leading to the 
fluorescence shown in 
Fig.~\ref{fig:spectra}(c)).

Figure~\ref{fig:spectra}(b)
shows resolved radiative decays 
from the 
$B\,^2\Sigma_{1/2}(v$$=$$0)$
state
to the 
$X\,^2\Sigma^+_{1/2}(v$$=$$1)$
and
$X\,^2\Sigma^+_{1/2}(v$$=$$2)$
states.
The ratio of the
size of these peaks is 
consistent with the 
calculated 
\cite{hao2019high}
branching
ratios for
a free
BaF molecule
(of
18\%,
and
2\%).
The decay down to the 
$X\,^2\Sigma^+_{1/2}(v$$=$$0)$
state is not observed as 
it 
is too close to the 
excitation wavelength
to allow it to be resolved 
from scattered laser light.
Using the 
calculated
\cite{hao2019high}
branching ratios
of
80\%,
to 
the
$v$$=$$0$
state,
it can be inferred 
that this decay rate 
would be 
4.4~times
larger than the 
observed fluorescence
to the 
$v$$=$$1$
state.
Including the efficiency of 
photon detection, 
the estimated rate of photon
emission is 
$2.3\times10^{12}$
per second.
The total rate for decay 
through the 
$A^\prime\,^2\Delta_{3/2}$
state 
is 
$1.7\times10^{12}$
per second.

% Including this inferred
% $v$$=$$0$
% decay,  
% 10\%
% of the 
% observed
% radiative decay
% comes directly from the 
% $B\,^2\Sigma_{1/2}(v$$=$$0)$
% state,
% with the remaining 
% 90\%
% going through the  
% $A^\prime\,^2\Delta_{3/2}$
% state.

\begin{figure}
\includegraphics[width=3.2in]
{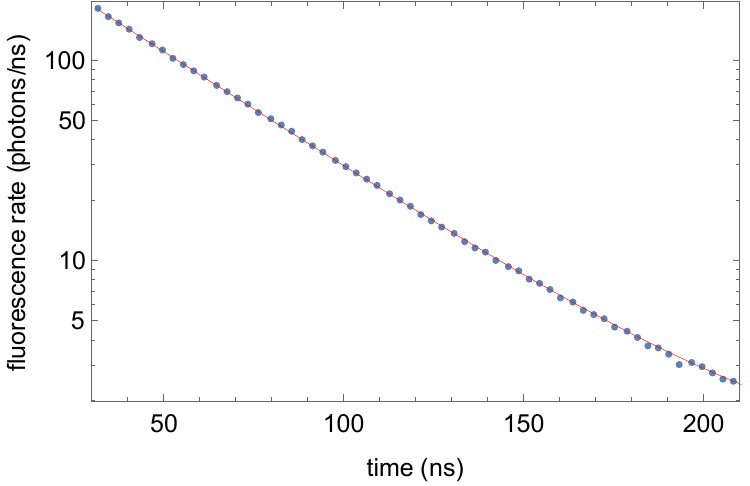}
\caption{Fluorescence from
decay from  
$B\,^2\Sigma_{1/2}(v$$=$$0)$
to  
$X\,^2\Sigma_{1/2}(v$$=$1)
after the excitation laser
is turned off at time
$t$$=$$0$,
presented on a logarithmic scale.
The solid line shows a fit
which gives a lifetime of 
37.5(3)~ns.
}
\label{fig:BstateLifetime}
\end{figure}

The lifetime of the 
$B\,^2\Sigma_{1/2}(v$$=$$0)$
state
in the matrix 
is measured by turning off the 
laser power 
(with an 
\mbox{acoustooptic}
modulator)
and observing 
(using repeated trials)
the 
time-resolved 
fluorescence
with a 
single-photon
counter.
The resulting fluorescence
is shown in 
Fig.~\ref{fig:BstateLifetime},
and a fit to this data leads to a measured 
lifetime 
of the 
$B\,^2\Sigma_{1/2}$
state
(within the matrix)
of 
37.5(3)~ns.
This value is close to the 
measured value
\cite{berg1993lifetime}
of 
41.7(3)~ns
and the calculated value
\cite{hao2019high}
of 
37.0~ns
for the free
BaF
molecule.

\begin{figure}
\includegraphics[width=3.2in]
{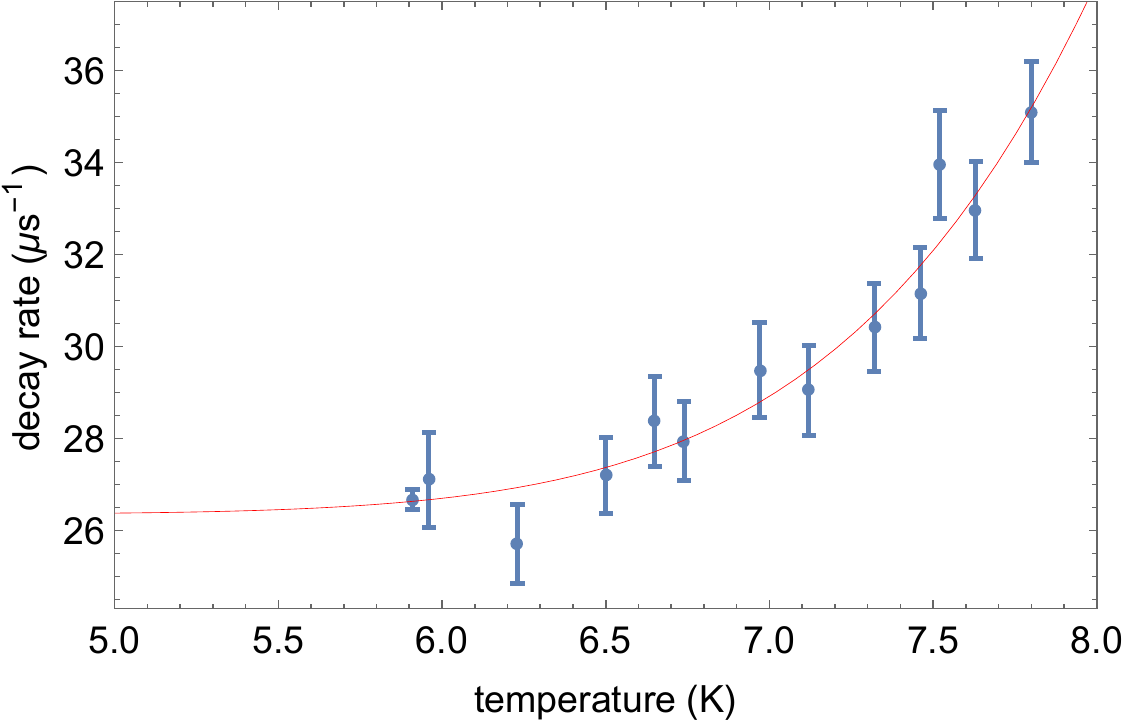}
\caption{The temperature 
dependence of the decay rate
from the 
$B\,^2\Sigma_{1/2}(v$$=$$0)$
state
is observed by repeating
measurements similar to that 
of 
Fig.~\ref{fig:BstateLifetime}
at elevated temperatures.
The fit line is to the 
Arrhenius form:
$
\tau_{\rm r}^{-1}
+
R_{\rm nr}
e^{-E_{\rm a}/(k_{\rm B} T)}
$,
with 
$E_{\rm a}$$=$59(11)~cm$^{-1}$.
The fact that the rate is 
almost
unchanging at the lower temperatures
indicates that 
radiative
decay dominates at temperatures below
6~K.
}
\label{fig:BstateDecayRateVsTemp}
\end{figure}

It is possible that some of the 
excited molecules return to the 
ground state without the emission
of a photon, 
giving the excess energy 
directly to the solid.
To test for such nonradiative decays, 
the decay rates are measured
with the temperature of the 
solid increased,
as shown in 
Fig.~\ref{fig:BstateDecayRateVsTemp}. 
The figure shows that the decay
rate increases with  
temperature
as
$
\tau_{\rm r}^{-1}
+
R_{\rm nr}
e^{-E_{\rm a}/(k_{\rm B} T)}
$,
where 
$\tau_{\rm r}$
is the 
radiative 
lifetime
and the constant
$R_{\rm nr}$
and the 
activation energy
$E_{\rm a}$
give the 
temperature-dependent
Arrhenius
expression
\cite{ramsthaler1986radiative}
for 
nonradiative 
decay.
At temperatures below
6~K,
the decay rate shows very 
little 
dependence on temperature,
indicating that 
almost all of the decay is 
radiative 
for our 
5.8-K
solid.

% lowest energies: 
%  0,
%  10$\,$718, 
%  11$\,$131,
%  11$\,$630, 
%  12$\,$262, 
%  14$\,$040, 
%  19$\,$988, 
%  20$\,$185. 

\section{saturation}

The rate equation 
that describes the 
laser excitation and 
decay 
(Fig.~\ref{fig:spectra}(a)
and
(b))
assuming a closed 
$X\,^2\Sigma^+_{1/2}$
$\leftrightarrow$
$B\,^2\Sigma_{1/2}$
system 
is
\begin{eqnarray}
\dot{b}
&=&
% \beta 
r g - b/\tau_b
=
% \beta 
r 
(
1
-b
% -a
)
- b/\tau_b,
% \nonumber
% \\
% \dot{a}
% &=&
% (1-\beta) r (1-b-a) - a/\tau_a,  
\label{eq:simpleRate}    
\end{eqnarray}
where
% $a$
% is the population in the
% $A^\prime\,^2\Delta_{3/2}$
% state,
$b$
is the population in the
$B\,^2\Sigma_{1/2}$
state
(at its equilibrium position),
$
1
-b
% -a
$
is the population 
in the ground state,
and
$r$
is the rate at which 
the 
ground-state
population
is excited to the 
$B\,^2\Sigma_{1/2}$
state.
This excitation rate is small
because of the short
time scale for the 
equilibration
of the nuclear positions
after the excitation.
It is proportional to laser
intensity,
and depends only on the 
ground-state 
population
(since the fast 
equilibration
leaves the state to which the 
molecules are excited 
essentially empty).
The solution to 
Eq.~(\ref{eq:simpleRate})
can be 
integrated over 
the
gaussian-beam 
profile
of the laser
\begin{equation}
r(s)
=
R \ e^{-\frac{s^2}{2\sigma^2}},   
\label{eq:rOfs}
\end{equation}
where 
$R$
is the excitation rate at the centre
of the laser beam
and 
$s$
is the distance from the central 
axis of the laser beam. 
The result gives
a 
radiative 
decay 
rate from the 
$B\,^2\Sigma_{1/2}$
state
of
\begin{eqnarray}
d(t)\!
=\!
\frac{b(t)}{\tau_b}
\!&=&\!
d_{\rm ss}
-
% r 
2 \pi \sigma^2 R
% \beta 
e^{-t/\tau_b}
+
% r^2 
\pi \sigma^2 R^2
% \beta
% \Big[
% e^{-t/\tau_a}
% \frac
% {\alpha \tau_a^2}
% {\tau_a\!-\!\tau_b}
% \nonumber
% \\
% &+&
(
t 
% \beta
% -
+
\tau_b
% \frac
% {\tau_a - 2 \beta \tau_a + \beta \tau_b}
% {\tau_a-\tau_b}
)
e^{-t/\tau_b}
% \Big]
\nonumber
\\
&+&
\mathcal{O}
(R^3 \tau_b^2),
\label{eq:timeProfile}
\end{eqnarray}
where
\begin{equation}
d_{\rm ss}
=
% r 
\pi \sigma^2 
% \beta
[
2 R
-
% r 
% \frac{1}{2} R
R^2
% (\alpha \tau_a
% +
% \beta 
\tau_b
% )
]  
\label{eq:steadyState}
\end{equation}
is the 
steady-state
value of 
$d(t)$.

We observe a nearly proportional 
increase in fluorescence as 
the laser intensity 
is increased
up to our maximum  
value
of 
80~kW/cm$^2$.
The small departure from 
proportionality 
(i.e.,
the small
terms that scale as 
$R^2$ 
that are indicating the 
onset of saturation)
of 
% Eqs.~(\ref{eq:steadyState})
% and
Eqs.~(\ref{eq:timeProfile})
and
(\ref{eq:steadyState})
can be used to estimate  
the value of 
$R$.
This procedure could easily 
make an overestimate of 
$R$,
as the excitation laser 
can cause temperature increases
(of the whole solid or 
locally near the molecules
being excited) 
and this increase
can also cause reductions
in fluorescence
(as seen,
e.g.,
in 
Fig.~\ref{fig:BstateDecayRateVsTemp}).
Thus,
values of 
$R$
obtained in this way 
should be considered to be 
upper limits on the 
actual value of 
$R$.

\begin{figure}
\includegraphics[width=3in]
{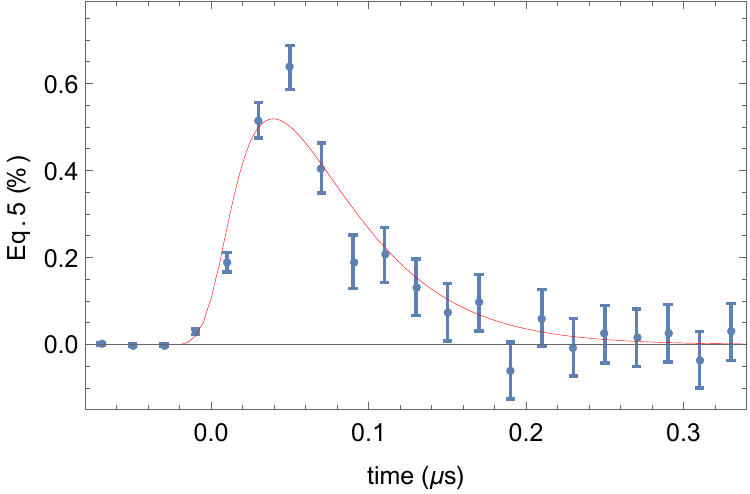}
\caption{
The time profile of the normalized difference 
of fluorescence at full and 
half power 
(see 
Eq.~(\ref{eq:fullHalf}))
reveals the excess fluorescence
that occurs before population 
builds up in the 
$B\,^2\Sigma_{1/2}$
state. 
The fit shown is to 
one parameter:
the laser excitation rate 
$R$.
}
\label{fig:fullHalfTimeProfile}
\end{figure}

Using 
Eq.~(\ref{eq:timeProfile}), 
the 
% steady-state fluorescence 
% increases by a factor of 
% 1.72 
% when the power is increased
% by a factor of 
% 1.95
% from a power 
% of 
% 198~mW
% to 
% a power
% of 
% 387~mW
% (for a 
% 40-micron
% 1/e$^2$-diameter
% laser beam).
% The implied value of 
% $R$ 
% is
% 0.86(1)~$\mu$s$^{-1}$
% per
% kW/cm$^2$.
% The
% ratio of the observed 
% steady-state
% decay at two powers
% ,
% using
% \begin{equation}
% \int_0^\infty
% r^N(s)
% s\,ds
% =
% \frac{2}{N}
% \pi \sigma^2
% R^N.
% \label{eq:gaussianIntegral}
% \end{equation}
% The 
normalized difference
of 
$d(t)$ 
at a power  
$P_0$$=$387~mW
(a rate 
$R_0$)
and at a reduced power
$\eta P_0$$=$$0.52 P_0$
leaves 
a residual
(to lowest order
in 
$R$)
of
\begin{eqnarray}
&&
\frac
{d(t)}
{d_{\rm ss}}
\Big|_{R_0}
-
\frac
{d(t)}
{d_{\rm ss}}
\Big|_{\eta R_0}
=
\frac{R_0}{2}
% \frac
% {
% \tau_a^2
(1-\eta)
% \alpha
% }
% {\tau_a-\tau_b}
% \Big[
% e^{-t/\tau_a}
% \nonumber
% \\
% &&
% -
e^{-t/\tau_b}
% \big(
% 1
% +
% \beta
t
% \frac
% {\tau_b-\tau_a}
% {
% \alpha
% \tau_a^2
% }
% \big)
% \Big]
.
\label{eq:fullHalf}
\end{eqnarray}
The observed difference
is shown in 
Fig.~\ref{fig:fullHalfTimeProfile}.
It agrees with the form of 
Eq.~(\ref{eq:fullHalf})
(convoluted with the 
25-ns
turn-on profile of the
laser),
as shown by the 
one-parameter 
fit 
(the solid line).
The fit value for 
$R_0$
implies that 
the upper limit of 
$R$
is
0.026(3)~$\mu$s$^{-1}$
per 
kW/cm$^2$.
A similar estimate 
results from looking at the 
saturation curve for the 
steady-state
fluorescence of
Eq.~(\ref{eq:steadyState}).
The estimate of 
$R$, 
along with the number of emitted 
photons,
implies that the 
lower limit of the number of 
BaF 
molecules
that are undergoing this
excitation and decay cycle 
is
$2\times10^9$
per 
mm$^2$.

A similar analysis for decay 
through the 
$A^\prime\,^2\Delta_{3/2}$
state
obtains an upper limit of 
0.004~$\mu$s$^{-1}$
per 
kW/cm$^2$.
This estimate is 
less precise
(approximately a 
50\%
uncertainty)
since the broad 
fluorescence peak of 
Fig.~\ref{fig:spectra}(c)
results in a range of lifetimes
(in the range of 
100
to 
1000~ns),
with the lifetimes varying 
over the wavelength 
of the fluorescence.
The implied number of 
BaF 
molecules undergoing 
the decay cycle through
the 
$A^\prime\,^2\Delta_{3/2}$
state
is  
greater than
$1.3\times10^{10}$
per 
mm$^2$
(with the same uncertainty of 
approximately
50\%). 

\section{conclusions}

A system for producing a 
neon solid doped with 
BaF 
molecules is 
described.
Laser-induced 
fluorescence of this
solid shows that 
there are
approximately
$10^{10}$
% or more
$^{138}$BaF
molecules
per
mm$^3$
present
in our solid that 
can be addressed 
via optical excitation.
This number is a large fraction of 
the
$5\times10^{10}$
$^{138}$BaF
molecules 
per 
mm$^2$
that 
arrive at the solid,
indicating that 
the solid is efficient
at isolating the 
BaF 
molecules
and that a large 
fraction of the molecules
are 
observable via optical 
excitation.
The 
$10^{10}$
addressable molecules
per
mm$^2$
is
the target 
value needed for the 
EDM$^3$
collaboration
to proceed towards 
an
eEDM 
measurement.
The number is many orders of 
magnitude larger
than the number of molecules
present at any time for 
other 
eEDM
experiments,
which use  
ion traps
\cite{Cairncross2017,
roussy2023new}
or
atomic beams
\cite{hudson2011improved,
baron2013order,
acme2018improved}.
Many steps still need 
to be demonstrated before
such a sample of 
BaF
molecules
can 
be used for an 
eEDM 
measurement, 
and these
steps will be addressed in 
a series of upcoming publications.

\vspace{7mm}
\section*{Acknowledgements}
This work is supported by the 
Alfred P. Sloan Foundation,
the 
Gordon and Betty Moore Foundation, 
the 
Templeton Foundation
in conjunction with the 
Northwestern Center for Fundamental Physics, 
the 
Canada Foundation for Innovation,
the
Ontario Research Fund,
the 
Natural Sciences and Engineering Research Council
of Canada,
and 
York University. 
Computations for this work were supported by 
Compute Ontario 
and 
the 
Digital Research Alliance of Canada. 
The authors are grateful for 
extensive discussions with 
Amar Vutha, 
Jaideep Singh,
and
René Fournier
which helped to guide this work.

\bibliography{BaF}

%merlin.mbs apsrev4-1.bst 2010-07-25 4.21a (PWD, AO, DPC) hacked
%Control: key (0)
%Control: author (0) dotless jnrlst
%Control: editor formatted (1) identically to author
%Control: production of article title (0) allowed
%Control: page (1) range
%Control: year (0) verbatim
%Control: production of eprint (0) enabled
\begin{thebibliography}{99}%
\makeatletter
\providecommand \@ifxundefined [1]{%
 \@ifx{#1\undefined}
}%
\providecommand \@ifnum [1]{%
 \ifnum #1\expandafter \@firstoftwo
 \else \expandafter \@secondoftwo
 \fi
}%
\providecommand \@ifx [1]{%
 \ifx #1\expandafter \@firstoftwo
 \else \expandafter \@secondoftwo
 \fi
}%
\providecommand \natexlab [1]{#1}%
\providecommand \enquote  [1]{``#1''}%
\providecommand \bibnamefont  [1]{#1}%
\providecommand \bibfnamefont [1]{#1}%
\providecommand \citenamefont [1]{#1}%
\providecommand \href@noop [0]{\@secondoftwo}%
\providecommand \href [0]{\begingroup \@sanitize@url \@href}%
\providecommand \@href[1]{\@@startlink{#1}\@@href}%
\providecommand \@@href[1]{\endgroup#1\@@endlink}%
\providecommand \@sanitize@url [0]{\catcode `\\12\catcode `\$12\catcode `\&12\catcode `\#12\catcode `\^12\catcode `\_12\catcode `\%12\relax}%
\providecommand \@@startlink[1]{}%
\providecommand \@@endlink[0]{}%
\providecommand \url  [0]{\begingroup\@sanitize@url \@url }%
\providecommand \@url [1]{\endgroup\@href {#1}{\urlprefix }}%
\providecommand \urlprefix  [0]{URL }%
\providecommand \Eprint [0]{\href }%
\providecommand \doibase [0]{http://dx.doi.org/}%
\providecommand \selectlanguage [0]{\@gobble}%
\providecommand \bibinfo  [0]{\@secondoftwo}%
\providecommand \bibfield  [0]{\@secondoftwo}%
\providecommand \translation [1]{[#1]}%
\providecommand \BibitemOpen [0]{}%
\providecommand \bibitemStop [0]{}%
\providecommand \bibitemNoStop [0]{.\EOS\space}%
\providecommand \EOS [0]{\spacefactor3000\relax}%
\providecommand \BibitemShut  [1]{\csname bibitem#1\endcsname}%
\let\auto@bib@innerbib\@empty
%</preamble>
\bibitem [{\citenamefont {Walters}\ and\ \citenamefont {Barratt}(1928)}]{walters1928alkaline}%
  \BibitemOpen
  \bibfield  {author} {\bibinfo {author} {\bibfnamefont {O.~H.}\ \bibnamefont {Walters}}\ and\ \bibinfo {author} {\bibfnamefont {S.}~\bibnamefont {Barratt}},\ }\bibfield  {title} {\enquote {\bibinfo {title} {{The alkaline earth halide spectra and their origin}},}\ }\href@noop {} {\bibfield  {journal} {\bibinfo  {journal} {Proceedings of the Royal Society of London. Series A}\ }\textbf {\bibinfo {volume} {118}},\ \bibinfo {pages} {120} (\bibinfo {year} {1928})}\BibitemShut {NoStop}%
\bibitem [{\citenamefont {Johnson}(1929)}]{Johnson1929band}%
  \BibitemOpen
  \bibfield  {author} {\bibinfo {author} {\bibfnamefont {R.~C.}\ \bibnamefont {Johnson}},\ }\bibfield  {title} {\enquote {\bibinfo {title} {{The band spectra of the alkaline earth halides. II. BaF, MgF}},}\ }\href@noop {} {\bibfield  {journal} {\bibinfo  {journal} {Proceedings of the Royal Society of London. Series A}\ }\textbf {\bibinfo {volume} {122}},\ \bibinfo {pages} {189} (\bibinfo {year} {1929})}\BibitemShut {NoStop}%
\bibitem [{\citenamefont {Nevin}(1931)}]{nevin1931spectrum}%
  \BibitemOpen
  \bibfield  {author} {\bibinfo {author} {\bibfnamefont {T.~E.}\ \bibnamefont {Nevin}},\ }\bibfield  {title} {\enquote {\bibinfo {title} {{The spectrum of barium fluoride in the extreme red and near infra-red}},}\ }\href@noop {} {\bibfield  {journal} {\bibinfo  {journal} {Proceedings of the Physical Society}\ }\textbf {\bibinfo {volume} {43}},\ \bibinfo {pages} {554} (\bibinfo {year} {1931})}\BibitemShut {NoStop}%
\bibitem [{\citenamefont {Jenkins}\ and\ \citenamefont {Harvey}(1932)}]{jenkins1932emission}%
  \BibitemOpen
  \bibfield  {author} {\bibinfo {author} {\bibfnamefont {F.~A.}\ \bibnamefont {Jenkins}}\ and\ \bibinfo {author} {\bibfnamefont {A.}~\bibnamefont {Harvey}},\ }\bibfield  {title} {\enquote {\bibinfo {title} {{Emission and absorption spectra of BaF}},}\ }\href@noop {} {\bibfield  {journal} {\bibinfo  {journal} {Physical Review}\ }\textbf {\bibinfo {volume} {39}},\ \bibinfo {pages} {922} (\bibinfo {year} {1932})}\BibitemShut {NoStop}%
\bibitem [{\citenamefont {Fowler~Jr}(1941)}]{fowler1941new}%
  \BibitemOpen
  \bibfield  {author} {\bibinfo {author} {\bibfnamefont {C.~A}\ \bibnamefont {Fowler~Jr}},\ }\bibfield  {title} {\enquote {\bibinfo {title} {{New absorption spectra of the alkaline earth fluorides}},}\ }\href@noop {} {\bibfield  {journal} {\bibinfo  {journal} {Physical Review}\ }\textbf {\bibinfo {volume} {59}},\ \bibinfo {pages} {645} (\bibinfo {year} {1941})}\BibitemShut {NoStop}%
\bibitem [{\citenamefont {Ehlert}\ \emph {et~al.}(1964)\citenamefont {Ehlert}, \citenamefont {Blue}, \citenamefont {Green},\ and\ \citenamefont {Margrave}}]{ehlert1964mass}%
  \BibitemOpen
  \bibfield  {author} {\bibinfo {author} {\bibfnamefont {T.~C.}\ \bibnamefont {Ehlert}}, \bibinfo {author} {\bibfnamefont {G.~D.}\ \bibnamefont {Blue}}, \bibinfo {author} {\bibfnamefont {J.~W.}\ \bibnamefont {Green}}, \ and\ \bibinfo {author} {\bibfnamefont {J.~L.}\ \bibnamefont {Margrave}},\ }\bibfield  {title} {\enquote {\bibinfo {title} {{Mass spectrometric studies at high temperatures. III. Dissociation energies of the alkaline earth monofluorides}},}\ }\href@noop {} {\bibfield  {journal} {\bibinfo  {journal} {The Journal of Chemical Physics}\ }\textbf {\bibinfo {volume} {41}},\ \bibinfo {pages} {2250} (\bibinfo {year} {1964})}\BibitemShut {NoStop}%
\bibitem [{\citenamefont {Hildenbrand}(1968)}]{hildenbrand1968mass}%
  \BibitemOpen
  \bibfield  {author} {\bibinfo {author} {\bibfnamefont {D.~L.}\ \bibnamefont {Hildenbrand}},\ }\bibfield  {title} {\enquote {\bibinfo {title} {{Mass-spectrometric studies of bonding in the group IIA fluorides}},}\ }\href@noop {} {\bibfield  {journal} {\bibinfo  {journal} {The Journal of Chemical Physics}\ }\textbf {\bibinfo {volume} {48}},\ \bibinfo {pages} {3657} (\bibinfo {year} {1968})}\BibitemShut {NoStop}%
\bibitem [{\citenamefont {Knight~Jr}\ \emph {et~al.}(1971)\citenamefont {Knight~Jr}, \citenamefont {Easley}, \citenamefont {Weltner~Jr},\ and\ \citenamefont {Wilson}}]{knight1971hyperfine}%
  \BibitemOpen
  \bibfield  {author} {\bibinfo {author} {\bibfnamefont {L.~B.}\ \bibnamefont {Knight~Jr}}, \bibinfo {author} {\bibfnamefont {W.~C.}\ \bibnamefont {Easley}}, \bibinfo {author} {\bibfnamefont {W.}~\bibnamefont {Weltner~Jr}}, \ and\ \bibinfo {author} {\bibfnamefont {M.}~\bibnamefont {Wilson}},\ }\bibfield  {title} {\enquote {\bibinfo {title} {{Hyperfine interaction and chemical bonding in MgF, CaF, SrF, and BaF molecules}},}\ }\href@noop {} {\bibfield  {journal} {\bibinfo  {journal} {The Journal of Chemical Physics}\ }\textbf {\bibinfo {volume} {54}},\ \bibinfo {pages} {322} (\bibinfo {year} {1971})}\BibitemShut {NoStop}%
\bibitem [{\citenamefont {Kushawaha}\ \emph {et~al.}(1972)\citenamefont {Kushawaha}, \citenamefont {Asthana}, \citenamefont {Shanker},\ and\ \citenamefont {Pathak}}]{kushawaha1972green}%
  \BibitemOpen
  \bibfield  {author} {\bibinfo {author} {\bibfnamefont {V.~S.}\ \bibnamefont {Kushawaha}}, \bibinfo {author} {\bibfnamefont {B.~P.}\ \bibnamefont {Asthana}}, \bibinfo {author} {\bibfnamefont {R.}~\bibnamefont {Shanker}}, \ and\ \bibinfo {author} {\bibfnamefont {C.~M.}\ \bibnamefont {Pathak}},\ }\bibfield  {title} {\enquote {\bibinfo {title} {{The green band system of BaF}},}\ }\href@noop {} {\bibfield  {journal} {\bibinfo  {journal} {Spectroscopy Letters}\ }\textbf {\bibinfo {volume} {5}},\ \bibinfo {pages} {407--414} (\bibinfo {year} {1972})}\BibitemShut {NoStop}%
\bibitem [{\citenamefont {Kushawaha}(1973)}]{kushawaha1973c2pi}%
  \BibitemOpen
  \bibfield  {author} {\bibinfo {author} {\bibfnamefont {V.~S.}\ \bibnamefont {Kushawaha}},\ }\bibfield  {title} {\enquote {\bibinfo {title} {{The C$\,^2\Pi$ and the X$\,^2\Sigma$ states of BaF}},}\ }\href@noop {} {\bibfield  {journal} {\bibinfo  {journal} {Spectroscopy Letters}\ }\textbf {\bibinfo {volume} {6}},\ \bibinfo {pages} {633} (\bibinfo {year} {1973})}\BibitemShut {NoStop}%
\bibitem [{\citenamefont {Dagdigian}\ \emph {et~al.}(1974)\citenamefont {Dagdigian}, \citenamefont {Cruse},\ and\ \citenamefont {Zare}}]{dagdigian1974radiative}%
  \BibitemOpen
  \bibfield  {author} {\bibinfo {author} {\bibfnamefont {P.~J.}\ \bibnamefont {Dagdigian}}, \bibinfo {author} {\bibfnamefont {H.~W.}\ \bibnamefont {Cruse}}, \ and\ \bibinfo {author} {\bibfnamefont {R.~N.}\ \bibnamefont {Zare}},\ }\bibfield  {title} {\enquote {\bibinfo {title} {{Radiative lifetimes of the alkaline earth monohalides}},}\ }\href@noop {} {\bibfield  {journal} {\bibinfo  {journal} {The Journal of Chemical Physics}\ }\textbf {\bibinfo {volume} {60}},\ \bibinfo {pages} {2330} (\bibinfo {year} {1974})}\BibitemShut {NoStop}%
\bibitem [{\citenamefont {Ryzlewicz}\ and\ \citenamefont {T{\"o}rring}(1980)}]{ryzlewicz1980formation}%
  \BibitemOpen
  \bibfield  {author} {\bibinfo {author} {\bibfnamefont {C.}~\bibnamefont {Ryzlewicz}}\ and\ \bibinfo {author} {\bibfnamefont {T.}~\bibnamefont {T{\"o}rring}},\ }\bibfield  {title} {\enquote {\bibinfo {title} {{Formation and microwave spectrum of the $^2\Sigma$-radical barium-monofluoride}},}\ }\href@noop {} {\bibfield  {journal} {\bibinfo  {journal} {Chemical Physics}\ }\textbf {\bibinfo {volume} {51}},\ \bibinfo {pages} {329} (\bibinfo {year} {1980})}\BibitemShut {NoStop}%
\bibitem [{\citenamefont {Ip}\ \emph {et~al.}(1981)\citenamefont {Ip}, \citenamefont {Bernath},\ and\ \citenamefont {Field}}]{Ip1981OpticalOptical}%
  \BibitemOpen
  \bibfield  {author} {\bibinfo {author} {\bibfnamefont {P.~C.~F.}\ \bibnamefont {Ip}}, \bibinfo {author} {\bibfnamefont {P.~F.}\ \bibnamefont {Bernath}}, \ and\ \bibinfo {author} {\bibfnamefont {R.~W.}\ \bibnamefont {Field}},\ }\bibfield  {title} {\enquote {\bibinfo {title} {{Optical-optical double-resonance spectroscopy of BaF: The E$\,^2\Sigma^+$ and F$\,^2\Pi$ states}},}\ }\href@noop {} {\bibfield  {journal} {\bibinfo  {journal} {Journal of Molecular Spectroscopy}\ }\textbf {\bibinfo {volume} {89}},\ \bibinfo {pages} {53} (\bibinfo {year} {1981})}\BibitemShut {NoStop}%
\bibitem [{\citenamefont {Ryzlewicz}\ \emph {et~al.}(1982)\citenamefont {Ryzlewicz}, \citenamefont {Sch{\"u}tze-Pahlmann}, \citenamefont {Hoeft},\ and\ \citenamefont {T{\"o}rring}}]{ryzlewicz1982rotational}%
  \BibitemOpen
  \bibfield  {author} {\bibinfo {author} {\bibfnamefont {C.}~\bibnamefont {Ryzlewicz}}, \bibinfo {author} {\bibfnamefont {H.-U.}\ \bibnamefont {Sch{\"u}tze-Pahlmann}}, \bibinfo {author} {\bibfnamefont {J.}~\bibnamefont {Hoeft}}, \ and\ \bibinfo {author} {\bibfnamefont {T.}~\bibnamefont {T{\"o}rring}},\ }\bibfield  {title} {\enquote {\bibinfo {title} {{Rotational spectrum and hyperfine structure of the $^2\Sigma$ radicals BaF and BaCl}},}\ }\href@noop {} {\bibfield  {journal} {\bibinfo  {journal} {Chemical Physics}\ }\textbf {\bibinfo {volume} {71}},\ \bibinfo {pages} {389--399} (\bibinfo {year} {1982})}\BibitemShut {NoStop}%
\bibitem [{\citenamefont {Ernst}\ \emph {et~al.}(1986)\citenamefont {Ernst}, \citenamefont {K{\"a}ndler},\ and\ \citenamefont {T{\"o}rring}}]{ernst1986hyperfine}%
  \BibitemOpen
  \bibfield  {author} {\bibinfo {author} {\bibfnamefont {W.~E.}\ \bibnamefont {Ernst}}, \bibinfo {author} {\bibfnamefont {J.}~\bibnamefont {K{\"a}ndler}}, \ and\ \bibinfo {author} {\bibfnamefont {T.}~\bibnamefont {T{\"o}rring}},\ }\bibfield  {title} {\enquote {\bibinfo {title} {{Hyperfine structure and electric dipole moment of BaF X $^2\Sigma^+$}},}\ }\href@noop {} {\bibfield  {journal} {\bibinfo  {journal} {The Journal of Chemical Physics}\ }\textbf {\bibinfo {volume} {84}},\ \bibinfo {pages} {4769--4773} (\bibinfo {year} {1986})}\BibitemShut {NoStop}%
\bibitem [{\citenamefont {Effantin}\ \emph {et~al.}(1987)\citenamefont {Effantin}, \citenamefont {d'INCAN}, \citenamefont {Bernard}, \citenamefont {Fabre}, \citenamefont {Stringat}, \citenamefont {Verg{\`e}s},\ and\ \citenamefont {Barrow}}]{effantin1987laser}%
  \BibitemOpen
  \bibfield  {author} {\bibinfo {author} {\bibfnamefont {C.}~\bibnamefont {Effantin}}, \bibinfo {author} {\bibfnamefont {J.}~\bibnamefont {d'INCAN}}, \bibinfo {author} {\bibfnamefont {A.}~\bibnamefont {Bernard}}, \bibinfo {author} {\bibfnamefont {G.}~\bibnamefont {Fabre}}, \bibinfo {author} {\bibfnamefont {R.}~\bibnamefont {Stringat}}, \bibinfo {author} {\bibfnamefont {J.}~\bibnamefont {Verg{\`e}s}}, \ and\ \bibinfo {author} {\bibfnamefont {R.}~\bibnamefont {Barrow}},\ }\bibfield  {title} {\enquote {\bibinfo {title} {{Laser-induced fluorescence from C$\,^2\Pi$ to X$\,^2\Sigma^+$ and H$^\prime\,^2\Delta$ states of BaF analysed by Fourier transform spectroscopy}},}\ }\href@noop {} {\bibfield  {journal} {\bibinfo  {journal} {Journal de Physique Colloques}\ }\textbf {\bibinfo {volume} {48}},\ \bibinfo {pages} {C7--673} (\bibinfo {year} {1987})}\BibitemShut {NoStop}%
\bibitem [{\citenamefont {Barrow}\ \emph {et~al.}(1988)\citenamefont {Barrow}, \citenamefont {Bernard}, \citenamefont {Effantin}, \citenamefont {d'Incan}, \citenamefont {Fabre}, \citenamefont {El~Hachimi}, \citenamefont {Stringat},\ and\ \citenamefont {Verg{\`e}s}}]{barrow1988metastable}%
  \BibitemOpen
  \bibfield  {author} {\bibinfo {author} {\bibfnamefont {R.~F.}\ \bibnamefont {Barrow}}, \bibinfo {author} {\bibfnamefont {A.}~\bibnamefont {Bernard}}, \bibinfo {author} {\bibfnamefont {C.}~\bibnamefont {Effantin}}, \bibinfo {author} {\bibfnamefont {J.}~\bibnamefont {d'Incan}}, \bibinfo {author} {\bibfnamefont {G.}~\bibnamefont {Fabre}}, \bibinfo {author} {\bibfnamefont {A.}~\bibnamefont {El~Hachimi}}, \bibinfo {author} {\bibfnamefont {R.}~\bibnamefont {Stringat}}, \ and\ \bibinfo {author} {\bibfnamefont {J.}~\bibnamefont {Verg{\`e}s}},\ }\bibfield  {title} {\enquote {\bibinfo {title} {{The metastable A$^\prime\,^2\Delta$ state of BaF}},}\ }\href@noop {} {\bibfield  {journal} {\bibinfo  {journal} {Chemical Physics Letters}\ }\textbf {\bibinfo {volume} {147}},\ \bibinfo {pages} {535} (\bibinfo {year} {1988})}\BibitemShut {NoStop}%
\bibitem [{\citenamefont {Effantin}\ \emph {et~al.}(1990)\citenamefont {Effantin}, \citenamefont {Bernard}, \citenamefont {d'Incan}, \citenamefont {Wannous}, \citenamefont {Verg{\`e}s},\ and\ \citenamefont {Barrow}}]{effantin1990studiesI}%
  \BibitemOpen
  \bibfield  {author} {\bibinfo {author} {\bibfnamefont {C.}~\bibnamefont {Effantin}}, \bibinfo {author} {\bibfnamefont {A.}~\bibnamefont {Bernard}}, \bibinfo {author} {\bibfnamefont {J.}~\bibnamefont {d'Incan}}, \bibinfo {author} {\bibfnamefont {G.}~\bibnamefont {Wannous}}, \bibinfo {author} {\bibfnamefont {J.}~\bibnamefont {Verg{\`e}s}}, \ and\ \bibinfo {author} {\bibfnamefont {R.~F.}\ \bibnamefont {Barrow}},\ }\bibfield  {title} {\enquote {\bibinfo {title} {{Studies of the electronic states of the BaF molecule: Part I: Effective constants for seven states below 30 000 cm$^{-1}$}},}\ }\href@noop {} {\bibfield  {journal} {\bibinfo  {journal} {Molecular Physics}\ }\textbf {\bibinfo {volume} {70}},\ \bibinfo {pages} {735} (\bibinfo {year} {1990})}\BibitemShut {NoStop}%
\bibitem [{\citenamefont {Bernard}\ \emph {et~al.}(1990)\citenamefont {Bernard}, \citenamefont {Effantin}, \citenamefont {d'Incan}, \citenamefont {Verg{\`e}s},\ and\ \citenamefont {Barrow}}]{bernard1990studiesII}%
  \BibitemOpen
  \bibfield  {author} {\bibinfo {author} {\bibfnamefont {A.}~\bibnamefont {Bernard}}, \bibinfo {author} {\bibfnamefont {C.}~\bibnamefont {Effantin}}, \bibinfo {author} {\bibfnamefont {J.}~\bibnamefont {d'Incan}}, \bibinfo {author} {\bibfnamefont {J.}~\bibnamefont {Verg{\`e}s}}, \ and\ \bibinfo {author} {\bibfnamefont {R.~F.}\ \bibnamefont {Barrow}},\ }\bibfield  {title} {\enquote {\bibinfo {title} {{Studies of the electronic states of the BaF molecule: Part II: The 5d ($v=$ 0, 1, 2) states}},}\ }\href@noop {} {\bibfield  {journal} {\bibinfo  {journal} {Molecular Physics}\ }\textbf {\bibinfo {volume} {70}},\ \bibinfo {pages} {747} (\bibinfo {year} {1990})}\BibitemShut {NoStop}%
\bibitem [{\citenamefont {Bernard}\ \emph {et~al.}(1992)\citenamefont {Bernard}, \citenamefont {Effantin}, \citenamefont {Andrianavalona}, \citenamefont {Verg{\`e}s},\ and\ \citenamefont {Barrow}}]{bernard1992laser}%
  \BibitemOpen
  \bibfield  {author} {\bibinfo {author} {\bibfnamefont {A.}~\bibnamefont {Bernard}}, \bibinfo {author} {\bibfnamefont {C.}~\bibnamefont {Effantin}}, \bibinfo {author} {\bibfnamefont {E.}~\bibnamefont {Andrianavalona}}, \bibinfo {author} {\bibfnamefont {J.}~\bibnamefont {Verg{\`e}s}}, \ and\ \bibinfo {author} {\bibfnamefont {R.~F.}\ \bibnamefont {Barrow}},\ }\bibfield  {title} {\enquote {\bibinfo {title} {{Laser-induced fluorescence of BaF: Further results for six electronic states}},}\ }\href@noop {} {\bibfield  {journal} {\bibinfo  {journal} {Journal of Molecular Spectroscopy}\ }\textbf {\bibinfo {volume} {152}},\ \bibinfo {pages} {174} (\bibinfo {year} {1992})}\BibitemShut {NoStop}%
\bibitem [{\citenamefont {Berg}\ \emph {et~al.}(1993)\citenamefont {Berg}, \citenamefont {Olsson}, \citenamefont {Chanteloup}, \citenamefont {Hishikawa},\ and\ \citenamefont {Royen}}]{berg1993lifetime}%
  \BibitemOpen
  \bibfield  {author} {\bibinfo {author} {\bibfnamefont {L.-E.}\ \bibnamefont {Berg}}, \bibinfo {author} {\bibfnamefont {T.}~\bibnamefont {Olsson}}, \bibinfo {author} {\bibfnamefont {J.-C.}\ \bibnamefont {Chanteloup}}, \bibinfo {author} {\bibfnamefont {A.}~\bibnamefont {Hishikawa}}, \ and\ \bibinfo {author} {\bibfnamefont {P.}~\bibnamefont {Royen}},\ }\bibfield  {title} {\enquote {\bibinfo {title} {{Lifetime measurements of excited molecular states using a Ti:sapphire laser: Radiative lifetimes of the A$\,^2\Pi$, B$\,^2\Sigma^+$ and C$\,^2\Pi$ states of BaF}},}\ }\href@noop {} {\bibfield  {journal} {\bibinfo  {journal} {Molecular Physics}\ }\textbf {\bibinfo {volume} {79}},\ \bibinfo {pages} {721} (\bibinfo {year} {1993})}\BibitemShut {NoStop}%
\bibitem [{\citenamefont {Jakubek}\ and\ \citenamefont {Field}(1994)}]{jakubek1994core}%
  \BibitemOpen
  \bibfield  {author} {\bibinfo {author} {\bibfnamefont {Z.~J.}\ \bibnamefont {Jakubek}}\ and\ \bibinfo {author} {\bibfnamefont {R.~W.}\ \bibnamefont {Field}},\ }\bibfield  {title} {\enquote {\bibinfo {title} {{Core-penetrating Rydberg series of BaF: s$\sim$p$\sim$d$\sim$f supercomplexes}},}\ }\href@noop {} {\bibfield  {journal} {\bibinfo  {journal} {Physical Review Letters}\ }\textbf {\bibinfo {volume} {72}},\ \bibinfo {pages} {2167} (\bibinfo {year} {1994})}\BibitemShut {NoStop}%
\bibitem [{\citenamefont {Jakubek}\ \emph {et~al.}(1994)\citenamefont {Jakubek}, \citenamefont {Harris}, \citenamefont {Field}, \citenamefont {Gardner},\ and\ \citenamefont {Murad}}]{jakubek1994ionization}%
  \BibitemOpen
  \bibfield  {author} {\bibinfo {author} {\bibfnamefont {Z.~J.}\ \bibnamefont {Jakubek}}, \bibinfo {author} {\bibfnamefont {N.~A.}\ \bibnamefont {Harris}}, \bibinfo {author} {\bibfnamefont {R.~W.}\ \bibnamefont {Field}}, \bibinfo {author} {\bibfnamefont {J.~A.}\ \bibnamefont {Gardner}}, \ and\ \bibinfo {author} {\bibfnamefont {E.}~\bibnamefont {Murad}},\ }\bibfield  {title} {\enquote {\bibinfo {title} {{Ionization potentials of CaF and BaF}},}\ }\href@noop {} {\bibfield  {journal} {\bibinfo  {journal} {The Journal of chemical physics}\ }\textbf {\bibinfo {volume} {100}},\ \bibinfo {pages} {622} (\bibinfo {year} {1994})}\BibitemShut {NoStop}%
\bibitem [{\citenamefont {Guo}\ \emph {et~al.}(1995)\citenamefont {Guo}, \citenamefont {Zhang},\ and\ \citenamefont {Bernath}}]{guo1995high}%
  \BibitemOpen
  \bibfield  {author} {\bibinfo {author} {\bibfnamefont {B.}~\bibnamefont {Guo}}, \bibinfo {author} {\bibfnamefont {K.~Q.}\ \bibnamefont {Zhang}}, \ and\ \bibinfo {author} {\bibfnamefont {P.~F.}\ \bibnamefont {Bernath}},\ }\bibfield  {title} {\enquote {\bibinfo {title} {{High-resolution Fourier transform infrared emission spectra of barium monofluoride}},}\ }\href@noop {} {\bibfield  {journal} {\bibinfo  {journal} {Journal of Molecular Spectroscopy}\ }\textbf {\bibinfo {volume} {170}},\ \bibinfo {pages} {59--74} (\bibinfo {year} {1995})}\BibitemShut {NoStop}%
\bibitem [{\citenamefont {Jakubek}\ and\ \citenamefont {Field}(1996)}]{jakubek1996core}%
  \BibitemOpen
  \bibfield  {author} {\bibinfo {author} {\bibfnamefont {Z.~J.}\ \bibnamefont {Jakubek}}\ and\ \bibinfo {author} {\bibfnamefont {R.~W.}\ \bibnamefont {Field}},\ }\bibfield  {title} {\enquote {\bibinfo {title} {{Core-penetrating Rydberg series of BaF: New electronic states in the $n^*\sim4$ region}},}\ }\href@noop {} {\bibfield  {journal} {\bibinfo  {journal} {Journal of Molecular Spectroscopy}\ }\textbf {\bibinfo {volume} {179}},\ \bibinfo {pages} {99} (\bibinfo {year} {1996})}\BibitemShut {NoStop}%
\bibitem [{\citenamefont {Jakubek}\ and\ \citenamefont {Field}(1997)}]{jakubek1997rydberg}%
  \BibitemOpen
  \bibfield  {author} {\bibinfo {author} {\bibfnamefont {Z.~J.}\ \bibnamefont {Jakubek}}\ and\ \bibinfo {author} {\bibfnamefont {R.~W.}\ \bibnamefont {Field}},\ }\bibfield  {title} {\enquote {\bibinfo {title} {{Rydberg series of BaF: peturbation-facilitated studies of core-non-penetrating states}},}\ }\href@noop {} {\bibfield  {journal} {\bibinfo  {journal} {Philosophical Transactions of the Royal Society of London. Series A: Mathematical, Physical and Engineering Sciences}\ }\textbf {\bibinfo {volume} {355}},\ \bibinfo {pages} {1507} (\bibinfo {year} {1997})}\BibitemShut {NoStop}%
\bibitem [{\citenamefont {Berg}\ \emph {et~al.}(1998)\citenamefont {Berg}, \citenamefont {Gador}, \citenamefont {Husain}, \citenamefont {Ludwigs},\ and\ \citenamefont {Royen}}]{berg1998lifetime}%
  \BibitemOpen
  \bibfield  {author} {\bibinfo {author} {\bibfnamefont {L.-E.}\ \bibnamefont {Berg}}, \bibinfo {author} {\bibfnamefont {N.}~\bibnamefont {Gador}}, \bibinfo {author} {\bibfnamefont {D.}~\bibnamefont {Husain}}, \bibinfo {author} {\bibfnamefont {H.}~\bibnamefont {Ludwigs}}, \ and\ \bibinfo {author} {\bibfnamefont {P.}~\bibnamefont {Royen}},\ }\bibfield  {title} {\enquote {\bibinfo {title} {{Lifetime measurements of the A$\,^2\Pi_{1/2}$ state of BaF using laser spectroscopy}},}\ }\href@noop {} {\bibfield  {journal} {\bibinfo  {journal} {Chemical Physics Letters}\ }\textbf {\bibinfo {volume} {287}},\ \bibinfo {pages} {89} (\bibinfo {year} {1998})}\BibitemShut {NoStop}%
\bibitem [{\citenamefont {Jakubek}\ and\ \citenamefont {Field}(2001)}]{jakubek2001core}%
  \BibitemOpen
  \bibfield  {author} {\bibinfo {author} {\bibfnamefont {Z.~J.}\ \bibnamefont {Jakubek}}\ and\ \bibinfo {author} {\bibfnamefont {R.~W.}\ \bibnamefont {Field}},\ }\bibfield  {title} {\enquote {\bibinfo {title} {{Core-penetrating Rydberg series of BaF: single-state and two-state fits of new electronic states in the $4.4\le n^* \le 14.3$ region}},}\ }\href@noop {} {\bibfield  {journal} {\bibinfo  {journal} {Journal of Molecular Spectroscopy}\ }\textbf {\bibinfo {volume} {205}},\ \bibinfo {pages} {197} (\bibinfo {year} {2001})}\BibitemShut {NoStop}%
\bibitem [{\citenamefont {Steimle}\ \emph {et~al.}(2011)\citenamefont {Steimle}, \citenamefont {Frey}, \citenamefont {Le}, \citenamefont {DeMille}, \citenamefont {Rahmlow},\ and\ \citenamefont {Linton}}]{Steimle2011MolecularBeam}%
  \BibitemOpen
  \bibfield  {author} {\bibinfo {author} {\bibfnamefont {T.~C.}\ \bibnamefont {Steimle}}, \bibinfo {author} {\bibfnamefont {S.}~\bibnamefont {Frey}}, \bibinfo {author} {\bibfnamefont {A.}~\bibnamefont {Le}}, \bibinfo {author} {\bibfnamefont {D.}~\bibnamefont {DeMille}}, \bibinfo {author} {\bibfnamefont {D.~A.}\ \bibnamefont {Rahmlow}}, \ and\ \bibinfo {author} {\bibfnamefont {C.}~\bibnamefont {Linton}},\ }\bibfield  {title} {\enquote {\bibinfo {title} {{Molecular-beam optical Stark and Zeeman study of the A$\,^2\Pi$--X$\,^2\Sigma^+$ (0,0) band system of BaF}},}\ }\href@noop {} {\bibfield  {journal} {\bibinfo  {journal} {Physical Review A}\ }\textbf {\bibinfo {volume} {84}},\ \bibinfo {pages} {012508} (\bibinfo {year} {2011})}\BibitemShut {NoStop}%
\bibitem [{\citenamefont {Cahn}\ \emph {et~al.}(2014)\citenamefont {Cahn}, \citenamefont {Ammon}, \citenamefont {Kirilov}, \citenamefont {Gurevich}, \citenamefont {Murphree}, \citenamefont {Paolino}, \citenamefont {Rahmlow}, \citenamefont {Kozlov},\ and\ \citenamefont {DeMille}}]{cahn2014zeeman}%
  \BibitemOpen
  \bibfield  {author} {\bibinfo {author} {\bibfnamefont {S.~B.}\ \bibnamefont {Cahn}}, \bibinfo {author} {\bibfnamefont {J.}~\bibnamefont {Ammon}}, \bibinfo {author} {\bibfnamefont {E.}~\bibnamefont {Kirilov}}, \bibinfo {author} {\bibfnamefont {Y.~V.}\ \bibnamefont {Gurevich}}, \bibinfo {author} {\bibfnamefont {D.}~\bibnamefont {Murphree}}, \bibinfo {author} {\bibfnamefont {R.}~\bibnamefont {Paolino}}, \bibinfo {author} {\bibfnamefont {D.~A.}\ \bibnamefont {Rahmlow}}, \bibinfo {author} {\bibfnamefont {M.~G.}\ \bibnamefont {Kozlov}}, \ and\ \bibinfo {author} {\bibfnamefont {D.}~\bibnamefont {DeMille}},\ }\bibfield  {title} {\enquote {\bibinfo {title} {{Zeeman-tuned rotational level-crossing spectroscopy in a diatomic free radical}},}\ }\href@noop {} {\bibfield  {journal} {\bibinfo  {journal} {Physical Review Letters}\ }\textbf {\bibinfo {volume} {112}},\ \bibinfo {pages} {163002} (\bibinfo {year} {2014})}\BibitemShut {NoStop}%
\bibitem [{\citenamefont {Zhou}\ \emph {et~al.}(2015)\citenamefont {Zhou}, \citenamefont {Grimes}, \citenamefont {Barnum}, \citenamefont {Patterson}, \citenamefont {Coy}, \citenamefont {Klein}, \citenamefont {Muenter},\ and\ \citenamefont {Field}}]{zhou2015direct}%
  \BibitemOpen
  \bibfield  {author} {\bibinfo {author} {\bibfnamefont {Y.}~\bibnamefont {Zhou}}, \bibinfo {author} {\bibfnamefont {D.~D.}\ \bibnamefont {Grimes}}, \bibinfo {author} {\bibfnamefont {T.~J.}\ \bibnamefont {Barnum}}, \bibinfo {author} {\bibfnamefont {D.}~\bibnamefont {Patterson}}, \bibinfo {author} {\bibfnamefont {S.~L.}\ \bibnamefont {Coy}}, \bibinfo {author} {\bibfnamefont {E.}~\bibnamefont {Klein}}, \bibinfo {author} {\bibfnamefont {J.~S.}\ \bibnamefont {Muenter}}, \ and\ \bibinfo {author} {\bibfnamefont {R.~W.}\ \bibnamefont {Field}},\ }\bibfield  {title} {\enquote {\bibinfo {title} {{ Direct detection of Rydberg-Rydberg millimeter-wave transitions in a buffer gas cooled molecular beam}},}\ }\href@noop {} {\bibfield  {journal} {\bibinfo  {journal} {Chemical Physics Letters}\ }\textbf {\bibinfo {volume} {640}},\ \bibinfo {pages} {124} (\bibinfo {year} {2015})}\BibitemShut {NoStop}%
\bibitem [{\citenamefont {Bu}\ \emph {et~al.}(2017)\citenamefont {Bu}, \citenamefont {Chen}, \citenamefont {Lv},\ and\ \citenamefont {Yan}}]{Bu2017Cold}%
  \BibitemOpen
  \bibfield  {author} {\bibinfo {author} {\bibfnamefont {W.}~\bibnamefont {Bu}}, \bibinfo {author} {\bibfnamefont {T.}~\bibnamefont {Chen}}, \bibinfo {author} {\bibfnamefont {G.}~\bibnamefont {Lv}}, \ and\ \bibinfo {author} {\bibfnamefont {B.}~\bibnamefont {Yan}},\ }\bibfield  {title} {\enquote {\bibinfo {title} {{Cold collision and high-resolution spectroscopy of buffer-gas-cooled BaF molecules}},}\ }\href@noop {} {\bibfield  {journal} {\bibinfo  {journal} {Physical Review A}\ }\textbf {\bibinfo {volume} {95}},\ \bibinfo {pages} {032701} (\bibinfo {year} {2017})}\BibitemShut {NoStop}%
\bibitem [{\citenamefont {Cournol}\ \emph {et~al.}(2018)\citenamefont {Cournol}, \citenamefont {Pillet}, \citenamefont {Lignier},\ and\ \citenamefont {Comparat}}]{cournol2018rovibrational}%
  \BibitemOpen
  \bibfield  {author} {\bibinfo {author} {\bibfnamefont {A.}~\bibnamefont {Cournol}}, \bibinfo {author} {\bibfnamefont {P.}~\bibnamefont {Pillet}}, \bibinfo {author} {\bibfnamefont {H.}~\bibnamefont {Lignier}}, \ and\ \bibinfo {author} {\bibfnamefont {D.}~\bibnamefont {Comparat}},\ }\bibfield  {title} {\enquote {\bibinfo {title} {{Rovibrational optical pumping of a molecular beam}},}\ }\href@noop {} {\bibfield  {journal} {\bibinfo  {journal} {Physical Review A}\ }\textbf {\bibinfo {volume} {97}},\ \bibinfo {pages} {031401} (\bibinfo {year} {2018})}\BibitemShut {NoStop}%
\bibitem [{\citenamefont {Aggarwal}\ \emph {et~al.}(2019)\citenamefont {Aggarwal}, \citenamefont {Marshall}, \citenamefont {Bethlem}, \citenamefont {Boeschoten}, \citenamefont {Borschevsky}, \citenamefont {Denis}, \citenamefont {Esajas}, \citenamefont {Hao}, \citenamefont {Hoekstra}, \citenamefont {Jungmann}, \citenamefont {Meijknecht}, \citenamefont {Mooij}, \citenamefont {Timmermans}, \citenamefont {Touwen}, \citenamefont {Ubachs}, \citenamefont {Vermeulen}, \citenamefont {Willmann}, \citenamefont {Yin},\ and\ \citenamefont {Zapara}}]{aggarwal2019lifetime}%
  \BibitemOpen
  \bibfield  {author} {\bibinfo {author} {\bibfnamefont {P.}~\bibnamefont {Aggarwal}}, \bibinfo {author} {\bibfnamefont {V.~R.}\ \bibnamefont {Marshall}}, \bibinfo {author} {\bibfnamefont {H.~L.}\ \bibnamefont {Bethlem}}, \bibinfo {author} {\bibfnamefont {A.}~\bibnamefont {Boeschoten}}, \bibinfo {author} {\bibfnamefont {A.}~\bibnamefont {Borschevsky}}, \bibinfo {author} {\bibfnamefont {M.}~\bibnamefont {Denis}}, \bibinfo {author} {\bibfnamefont {K.}~\bibnamefont {Esajas}}, \bibinfo {author} {\bibfnamefont {Y.}~\bibnamefont {Hao}}, \bibinfo {author} {\bibfnamefont {S.}~\bibnamefont {Hoekstra}}, \bibinfo {author} {\bibfnamefont {K.}~\bibnamefont {Jungmann}}, \bibinfo {author} {\bibfnamefont {T.~B.}\ \bibnamefont {Meijknecht}}, \bibinfo {author} {\bibfnamefont {M.~C.}\ \bibnamefont {Mooij}}, \bibinfo {author} {\bibfnamefont {R.~G.~E.}\ \bibnamefont {Timmermans}}, \bibinfo {author} {\bibfnamefont {A.}~\bibnamefont {Touwen}}, \bibinfo {author} {\bibfnamefont {W.}~\bibnamefont {Ubachs}}, \bibinfo {author}
  {\bibfnamefont {S.~M.}\ \bibnamefont {Vermeulen}}, \bibinfo {author} {\bibfnamefont {L.}~\bibnamefont {Willmann}}, \bibinfo {author} {\bibfnamefont {Y.}~\bibnamefont {Yin}}, \ and\ \bibinfo {author} {\bibfnamefont {A.}~\bibnamefont {Zapara}},\ }\bibfield  {title} {\enquote {\bibinfo {title} {{Lifetime measurements of the A$\,^2\Pi_{1/2}$ and A$\,^2\Pi_{3/2}$ states in BaF}},}\ }\href@noop {} {\bibfield  {journal} {\bibinfo  {journal} {Physical Review A}\ }\textbf {\bibinfo {volume} {100}},\ \bibinfo {pages} {052503} (\bibinfo {year} {2019})}\BibitemShut {NoStop}%
\bibitem [{\citenamefont {Aggarwal}\ \emph {et~al.}(2021)\citenamefont {Aggarwal}, \citenamefont {Bethlem}, \citenamefont {Boeschoten}, \citenamefont {Borschevsky}, \citenamefont {Esajas}, \citenamefont {Hao}, \citenamefont {Hoekstra}, \citenamefont {Jungmann}, \citenamefont {Marshall}, \citenamefont {Meijknecht}, \citenamefont {Timmermans}, \citenamefont {Touwen}, \citenamefont {Ubachs}, \citenamefont {Willmann}, \citenamefont {Yin},\ and\ \citenamefont {Zapara}}]{aggarwal2021supersonic}%
  \BibitemOpen
  \bibfield  {author} {\bibinfo {author} {\bibfnamefont {P.}~\bibnamefont {Aggarwal}}, \bibinfo {author} {\bibfnamefont {H.~L.}\ \bibnamefont {Bethlem}}, \bibinfo {author} {\bibfnamefont {A.}~\bibnamefont {Boeschoten}}, \bibinfo {author} {\bibfnamefont {A.}~\bibnamefont {Borschevsky}}, \bibinfo {author} {\bibfnamefont {K.}~\bibnamefont {Esajas}}, \bibinfo {author} {\bibfnamefont {Y.}~\bibnamefont {Hao}}, \bibinfo {author} {\bibfnamefont {S.}~\bibnamefont {Hoekstra}}, \bibinfo {author} {\bibfnamefont {K.}~\bibnamefont {Jungmann}}, \bibinfo {author} {\bibfnamefont {V.~R.}\ \bibnamefont {Marshall}}, \bibinfo {author} {\bibfnamefont {T.~B.}\ \bibnamefont {Meijknecht}}, \bibinfo {author} {\bibfnamefont {R.~G.~E.}\ \bibnamefont {Timmermans}}, \bibinfo {author} {\bibfnamefont {A.}~\bibnamefont {Touwen}}, \bibinfo {author} {\bibfnamefont {W.}~\bibnamefont {Ubachs}}, \bibinfo {author} {\bibfnamefont {L.}~\bibnamefont {Willmann}}, \bibinfo {author} {\bibfnamefont {Y.}~\bibnamefont {Yin}}, \ and\ \bibinfo {author}
  {\bibfnamefont {A.}~\bibnamefont {Zapara}},\ }\bibfield  {title} {\enquote {\bibinfo {title} {{A supersonic laser ablation beam source with narrow velocity spreads}},}\ }\href@noop {} {\bibfield  {journal} {\bibinfo  {journal} {Review of Scientific Instruments}\ }\textbf {\bibinfo {volume} {92}} (\bibinfo {year} {2021})}\BibitemShut {NoStop}%
\bibitem [{\citenamefont {Bu}\ \emph {et~al.}(2022)\citenamefont {Bu}, \citenamefont {Zhang}, \citenamefont {Liang}, \citenamefont {Chen},\ and\ \citenamefont {Yan}}]{bu2022saturated}%
  \BibitemOpen
  \bibfield  {author} {\bibinfo {author} {\bibfnamefont {W.}~\bibnamefont {Bu}}, \bibinfo {author} {\bibfnamefont {Y.}~\bibnamefont {Zhang}}, \bibinfo {author} {\bibfnamefont {Q.}~\bibnamefont {Liang}}, \bibinfo {author} {\bibfnamefont {T.}~\bibnamefont {Chen}}, \ and\ \bibinfo {author} {\bibfnamefont {B.}~\bibnamefont {Yan}},\ }\bibfield  {title} {\enquote {\bibinfo {title} {{Saturated absorption spectroscopy of buffer-gas-cooled barium monofluoride molecules}},}\ }\href@noop {} {\bibfield  {journal} {\bibinfo  {journal} {Frontiers of Physics}\ }\textbf {\bibinfo {volume} {17}},\ \bibinfo {pages} {62502} (\bibinfo {year} {2022})}\BibitemShut {NoStop}%
\bibitem [{\citenamefont {Courageux}\ \emph {et~al.}(2022)\citenamefont {Courageux}, \citenamefont {Cournol}, \citenamefont {Comparat}, \citenamefont {Viaris~de Lesegno},\ and\ \citenamefont {Lignier}}]{courageux2022efficient}%
  \BibitemOpen
  \bibfield  {author} {\bibinfo {author} {\bibfnamefont {T.}~\bibnamefont {Courageux}}, \bibinfo {author} {\bibfnamefont {A.}~\bibnamefont {Cournol}}, \bibinfo {author} {\bibfnamefont {D.}~\bibnamefont {Comparat}}, \bibinfo {author} {\bibfnamefont {B.}~\bibnamefont {Viaris~de Lesegno}}, \ and\ \bibinfo {author} {\bibfnamefont {H.}~\bibnamefont {Lignier}},\ }\bibfield  {title} {\enquote {\bibinfo {title} {{Efficient rotational cooling of a cold beam of barium monofluoride}},}\ }\href@noop {} {\bibfield  {journal} {\bibinfo  {journal} {New Journal of Physics}\ }\textbf {\bibinfo {volume} {24}},\ \bibinfo {pages} {025007} (\bibinfo {year} {2022})}\BibitemShut {NoStop}%
\bibitem [{\citenamefont {Mooij}\ \emph {et~al.}(2024)\citenamefont {Mooij}, \citenamefont {Bethlem}, \citenamefont {Boeschoten}, \citenamefont {Borschevsky}, \citenamefont {Esajas}, \citenamefont {Fikkers}, \citenamefont {Hoekstra}, \citenamefont {van Hofslot}, \citenamefont {Jungmann}, \citenamefont {Marshall}, \citenamefont {Meijknecht}, \citenamefont {Timmermans}, \citenamefont {Touwen}, \citenamefont {Ubachs}, \citenamefont {Willmann},\ and\ \citenamefont {Yin}}]{mooij2024influence}%
  \BibitemOpen
  \bibfield  {author} {\bibinfo {author} {\bibfnamefont {M.~C.}\ \bibnamefont {Mooij}}, \bibinfo {author} {\bibfnamefont {H.~L.}\ \bibnamefont {Bethlem}}, \bibinfo {author} {\bibfnamefont {A.}~\bibnamefont {Boeschoten}}, \bibinfo {author} {\bibfnamefont {A.}~\bibnamefont {Borschevsky}}, \bibinfo {author} {\bibfnamefont {K.}~\bibnamefont {Esajas}}, \bibinfo {author} {\bibfnamefont {T.~H.}\ \bibnamefont {Fikkers}}, \bibinfo {author} {\bibfnamefont {S.}~\bibnamefont {Hoekstra}}, \bibinfo {author} {\bibfnamefont {J.~W.~F.}\ \bibnamefont {van Hofslot}}, \bibinfo {author} {\bibfnamefont {K.}~\bibnamefont {Jungmann}}, \bibinfo {author} {\bibfnamefont {V.~R.}\ \bibnamefont {Marshall}}, \bibinfo {author} {\bibfnamefont {T.~B.}\ \bibnamefont {Meijknecht}}, \bibinfo {author} {\bibfnamefont {R.~G.~E.}\ \bibnamefont {Timmermans}}, \bibinfo {author} {\bibfnamefont {A.}~\bibnamefont {Touwen}}, \bibinfo {author} {\bibfnamefont {W.}~\bibnamefont {Ubachs}}, \bibinfo {author} {\bibfnamefont {L.}~\bibnamefont {Willmann}}, \ and\
  \bibinfo {author} {\bibfnamefont {Y.}~\bibnamefont {Yin}},\ }\bibfield  {title} {\enquote {\bibinfo {title} {{Influence of source parameters on the longitudinal phase-space distribution of a pulsed cryogenic beam of barium fluoride molecules}},}\ }\href@noop {} {\bibfield  {journal} {\bibinfo  {journal} {arXiv preprint arXiv:2401.16590}\ } (\bibinfo {year} {2024})}\BibitemShut {NoStop}%
\bibitem [{\citenamefont {Touwen}\ \emph {et~al.}(2024)\citenamefont {Touwen}, \citenamefont {van Hofslot}, \citenamefont {Qualm}, \citenamefont {Borchers}, \citenamefont {Bause}, \citenamefont {Bethlem}, \citenamefont {Boeschoten}, \citenamefont {Borschevsky}, \citenamefont {Fikkers}, \citenamefont {Hoekstra}, \citenamefont {Jungmann}, \citenamefont {Marshall}, \citenamefont {Meijknecht}, \citenamefont {Mooij}, \citenamefont {Timmermans}, \citenamefont {Ubachs},\ and\ \citenamefont {Willmann}}]{touwen2024manipulating}%
  \BibitemOpen
  \bibfield  {author} {\bibinfo {author} {\bibfnamefont {A.}~\bibnamefont {Touwen}}, \bibinfo {author} {\bibfnamefont {J.~W.~F.}\ \bibnamefont {van Hofslot}}, \bibinfo {author} {\bibfnamefont {T.}~\bibnamefont {Qualm}}, \bibinfo {author} {\bibfnamefont {R.}~\bibnamefont {Borchers}}, \bibinfo {author} {\bibfnamefont {R.}~\bibnamefont {Bause}}, \bibinfo {author} {\bibfnamefont {H.~L.}\ \bibnamefont {Bethlem}}, \bibinfo {author} {\bibfnamefont {A.}~\bibnamefont {Boeschoten}}, \bibinfo {author} {\bibfnamefont {A.}~\bibnamefont {Borschevsky}}, \bibinfo {author} {\bibfnamefont {T.~H.}\ \bibnamefont {Fikkers}}, \bibinfo {author} {\bibfnamefont {S.}~\bibnamefont {Hoekstra}}, \bibinfo {author} {\bibfnamefont {K.}~\bibnamefont {Jungmann}}, \bibinfo {author} {\bibfnamefont {V.~R.}\ \bibnamefont {Marshall}}, \bibinfo {author} {\bibfnamefont {T.~B.}\ \bibnamefont {Meijknecht}}, \bibinfo {author} {\bibfnamefont {M.~C.}\ \bibnamefont {Mooij}}, \bibinfo {author} {\bibfnamefont {R.~G.~E.}\ \bibnamefont {Timmermans}}, \bibinfo
  {author} {\bibfnamefont {W.}~\bibnamefont {Ubachs}}, \ and\ \bibinfo {author} {\bibfnamefont {L.}~\bibnamefont {Willmann}},\ }\bibfield  {title} {\enquote {\bibinfo {title} {{Manipulating a beam of barium fluoride molecules using an electrostatic hexapole}},}\ }\href@noop {} {\bibfield  {journal} {\bibinfo  {journal} {arXiv preprint arXiv:2402.09300}\ } (\bibinfo {year} {2024})}\BibitemShut {NoStop}%
\bibitem [{\citenamefont {T{\"o}rring}\ \emph {et~al.}(1989)\citenamefont {T{\"o}rring}, \citenamefont {Ernst},\ and\ \citenamefont {K{\"a}ndler}}]{torring1989energies}%
  \BibitemOpen
  \bibfield  {author} {\bibinfo {author} {\bibfnamefont {T.}~\bibnamefont {T{\"o}rring}}, \bibinfo {author} {\bibfnamefont {W.~E.}\ \bibnamefont {Ernst}}, \ and\ \bibinfo {author} {\bibfnamefont {J.}~\bibnamefont {K{\"a}ndler}},\ }\bibfield  {title} {\enquote {\bibinfo {title} {{Energies and electric dipole moments of the low lying electronic states of the alkaline earth monohalides from an electrostatic polarization model}},}\ }\href@noop {} {\bibfield  {journal} {\bibinfo  {journal} {The Journal of Chemical Physics}\ }\textbf {\bibinfo {volume} {90}},\ \bibinfo {pages} {4927} (\bibinfo {year} {1989})}\BibitemShut {NoStop}%
\bibitem [{\citenamefont {Allouche}\ \emph {et~al.}(1993)\citenamefont {Allouche}, \citenamefont {Wannous},\ and\ \citenamefont {Aubert-Fr{\'e}con}}]{allouche1993ligand}%
  \BibitemOpen
  \bibfield  {author} {\bibinfo {author} {\bibfnamefont {A.~R.}\ \bibnamefont {Allouche}}, \bibinfo {author} {\bibfnamefont {G.}~\bibnamefont {Wannous}}, \ and\ \bibinfo {author} {\bibfnamefont {M.}~\bibnamefont {Aubert-Fr{\'e}con}},\ }\bibfield  {title} {\enquote {\bibinfo {title} {{A ligand-field approach for the low-lying states of Ca, Sr and Ba monohalides}},}\ }\href@noop {} {\bibfield  {journal} {\bibinfo  {journal} {Chemical physics}\ }\textbf {\bibinfo {volume} {170}},\ \bibinfo {pages} {11} (\bibinfo {year} {1993})}\BibitemShut {NoStop}%
\bibitem [{\citenamefont {Arif}\ \emph {et~al.}(1997)\citenamefont {Arif}, \citenamefont {Jungen},\ and\ \citenamefont {Roche}}]{arif1997rydberg}%
  \BibitemOpen
  \bibfield  {author} {\bibinfo {author} {\bibfnamefont {M.}~\bibnamefont {Arif}}, \bibinfo {author} {\bibfnamefont {C.}~\bibnamefont {Jungen}}, \ and\ \bibinfo {author} {\bibfnamefont {A.~L.}\ \bibnamefont {Roche}},\ }\bibfield  {title} {\enquote {\bibinfo {title} {{The Rydberg spectrum of CaF and BaF: Calculation by R-matrix and generalized quantum defect theory}},}\ }\href@noop {} {\bibfield  {journal} {\bibinfo  {journal} {The Journal of Chemical Physics}\ }\textbf {\bibinfo {volume} {106}},\ \bibinfo {pages} {4102} (\bibinfo {year} {1997})}\BibitemShut {NoStop}%
\bibitem [{\citenamefont {Kobus}\ \emph {et~al.}(2002)\citenamefont {Kobus}, \citenamefont {Moncrieff},\ and\ \citenamefont {Wilson}}]{kobus2002comparison}%
  \BibitemOpen
  \bibfield  {author} {\bibinfo {author} {\bibfnamefont {J.}~\bibnamefont {Kobus}}, \bibinfo {author} {\bibfnamefont {D.}~\bibnamefont {Moncrieff}}, \ and\ \bibinfo {author} {\bibfnamefont {S.}~\bibnamefont {Wilson}},\ }\bibfield  {title} {\enquote {\bibinfo {title} {{A comparison of finite basis set and finite difference Hartree—Fock calculations for the open-shell (X$\,^2\Sigma^+$) molecules BaF and YbF}},}\ }\href@noop {} {\bibfield  {journal} {\bibinfo  {journal} {Molecular Physics}\ }\textbf {\bibinfo {volume} {100}},\ \bibinfo {pages} {499--508} (\bibinfo {year} {2002})}\BibitemShut {NoStop}%
\bibitem [{\citenamefont {Tohme}\ and\ \citenamefont {Korek}(2015)}]{tohme2015theoretical}%
  \BibitemOpen
  \bibfield  {author} {\bibinfo {author} {\bibfnamefont {S.~N.}\ \bibnamefont {Tohme}}\ and\ \bibinfo {author} {\bibfnamefont {M.}~\bibnamefont {Korek}},\ }\bibfield  {title} {\enquote {\bibinfo {title} {{Theoretical study of the electronic structure with dipole moment calculations of barium monofluoride}},}\ }\href@noop {} {\bibfield  {journal} {\bibinfo  {journal} {Journal of Quantitative Spectroscopy and Radiative Transfer}\ }\textbf {\bibinfo {volume} {167}},\ \bibinfo {pages} {82} (\bibinfo {year} {2015})}\BibitemShut {NoStop}%
\bibitem [{\citenamefont {Prasannaa}\ \emph {et~al.}(2016)\citenamefont {Prasannaa}, \citenamefont {Sreerekha}, \citenamefont {Abe}, \citenamefont {Bannur},\ and\ \citenamefont {Das}}]{prasannaa2016permanent}%
  \BibitemOpen
  \bibfield  {author} {\bibinfo {author} {\bibfnamefont {V.~S.}\ \bibnamefont {Prasannaa}}, \bibinfo {author} {\bibfnamefont {S.}~\bibnamefont {Sreerekha}}, \bibinfo {author} {\bibfnamefont {M.}~\bibnamefont {Abe}}, \bibinfo {author} {\bibfnamefont {V.~M.}\ \bibnamefont {Bannur}}, \ and\ \bibinfo {author} {\bibfnamefont {B.~P.}\ \bibnamefont {Das}},\ }\bibfield  {title} {\enquote {\bibinfo {title} {{Permanent electric dipole moments of alkaline-earth-metal monofluorides: Interplay of relativistic and correlation effects}},}\ }\href@noop {} {\bibfield  {journal} {\bibinfo  {journal} {Physical Review A}\ }\textbf {\bibinfo {volume} {93}},\ \bibinfo {pages} {042504} (\bibinfo {year} {2016})}\BibitemShut {NoStop}%
\bibitem [{\citenamefont {Bala}\ \emph {et~al.}(2019)\citenamefont {Bala}, \citenamefont {Nataraj},\ and\ \citenamefont {Nayak}}]{bala2019ab}%
  \BibitemOpen
  \bibfield  {author} {\bibinfo {author} {\bibfnamefont {R.}~\bibnamefont {Bala}}, \bibinfo {author} {\bibfnamefont {H.~S.}\ \bibnamefont {Nataraj}}, \ and\ \bibinfo {author} {\bibfnamefont {M.~K.}\ \bibnamefont {Nayak}},\ }\bibfield  {title} {\enquote {\bibinfo {title} {{Ab initio calculations of permanent dipole moments and dipole polarizabilities of alkaline-earth monofluorides}},}\ }\href@noop {} {\bibfield  {journal} {\bibinfo  {journal} {Journal of Physics B: Atomic, Molecular and Optical Physics}\ }\textbf {\bibinfo {volume} {52}},\ \bibinfo {pages} {085101} (\bibinfo {year} {2019})}\BibitemShut {NoStop}%
\bibitem [{\citenamefont {Haase}\ \emph {et~al.}(2020)\citenamefont {Haase}, \citenamefont {Eliav}, \citenamefont {Ilia\v{s}},\ and\ \citenamefont {Borschevsky}}]{haase2020hyperfine}%
  \BibitemOpen
  \bibfield  {author} {\bibinfo {author} {\bibfnamefont {P.~A.~B.}\ \bibnamefont {Haase}}, \bibinfo {author} {\bibfnamefont {E.}~\bibnamefont {Eliav}}, \bibinfo {author} {\bibfnamefont {M.}~\bibnamefont {Ilia\v{s}}}, \ and\ \bibinfo {author} {\bibfnamefont {A.}~\bibnamefont {Borschevsky}},\ }\bibfield  {title} {\enquote {\bibinfo {title} {{Hyperfine structure constants on the relativistic coupled cluster level with associated uncertainties}},}\ }\href@noop {} {\bibfield  {journal} {\bibinfo  {journal} {The Journal of Physical Chemistry A}\ }\textbf {\bibinfo {volume} {124}},\ \bibinfo {pages} {3157} (\bibinfo {year} {2020})}\BibitemShut {NoStop}%
\bibitem [{\citenamefont {Skripnikov}\ \emph {et~al.}(2021)\citenamefont {Skripnikov}, \citenamefont {Chubukov},\ and\ \citenamefont {Shakhova}}]{skripnikov2021role}%
  \BibitemOpen
  \bibfield  {author} {\bibinfo {author} {\bibfnamefont {L.~V.}\ \bibnamefont {Skripnikov}}, \bibinfo {author} {\bibfnamefont {D.~V.}\ \bibnamefont {Chubukov}}, \ and\ \bibinfo {author} {\bibfnamefont {V.~M.}\ \bibnamefont {Shakhova}},\ }\bibfield  {title} {\enquote {\bibinfo {title} {{The role of QED effects in transition energies of heavy-atom alkaline earth monofluoride molecules: A theoretical study of Ba$^+$, BaF, RaF, and E120F}},}\ }\href@noop {} {\bibfield  {journal} {\bibinfo  {journal} {The Journal of Chemical Physics}\ }\textbf {\bibinfo {volume} {155}} (\bibinfo {year} {2021})}\BibitemShut {NoStop}%
\bibitem [{\citenamefont {Denis}\ \emph {et~al.}(2022)\citenamefont {Denis}, \citenamefont {Haase}, \citenamefont {Mooij}, \citenamefont {Chamorro}, \citenamefont {Aggarwal}, \citenamefont {Bethlem}, \citenamefont {Boeschoten}, \citenamefont {Borschevsky}, \citenamefont {Esajas}, \citenamefont {Hao}, \citenamefont {Hoekstra}, \citenamefont {van Hofslot}, \citenamefont {Marshall}, \citenamefont {Meijknecht}, \citenamefont {Timmermans}, \citenamefont {Touwen}, \citenamefont {Ubachs}, \citenamefont {Willmann},\ and\ \citenamefont {Y.}}]{denis2022benchmarking}%
  \BibitemOpen
  \bibfield  {author} {\bibinfo {author} {\bibfnamefont {M.}~\bibnamefont {Denis}}, \bibinfo {author} {\bibfnamefont {P.~A.~B.}\ \bibnamefont {Haase}}, \bibinfo {author} {\bibfnamefont {M.~C.}\ \bibnamefont {Mooij}}, \bibinfo {author} {\bibfnamefont {Y.}~\bibnamefont {Chamorro}}, \bibinfo {author} {\bibfnamefont {P.}~\bibnamefont {Aggarwal}}, \bibinfo {author} {\bibfnamefont {H.~L.}\ \bibnamefont {Bethlem}}, \bibinfo {author} {\bibfnamefont {A.}~\bibnamefont {Boeschoten}}, \bibinfo {author} {\bibfnamefont {A.}~\bibnamefont {Borschevsky}}, \bibinfo {author} {\bibfnamefont {K.}~\bibnamefont {Esajas}}, \bibinfo {author} {\bibfnamefont {Y.}~\bibnamefont {Hao}}, \bibinfo {author} {\bibfnamefont {S.}~\bibnamefont {Hoekstra}}, \bibinfo {author} {\bibfnamefont {J.~W.~F.}\ \bibnamefont {van Hofslot}}, \bibinfo {author} {\bibfnamefont {V.~R.}\ \bibnamefont {Marshall}}, \bibinfo {author} {\bibfnamefont {T.~B.}\ \bibnamefont {Meijknecht}}, \bibinfo {author} {\bibfnamefont {R.~G.~E.}\ \bibnamefont {Timmermans}}, \bibinfo
  {author} {\bibfnamefont {A.}~\bibnamefont {Touwen}}, \bibinfo {author} {\bibfnamefont {W.}~\bibnamefont {Ubachs}}, \bibinfo {author} {\bibfnamefont {L.}~\bibnamefont {Willmann}}, \ and\ \bibinfo {author} {\bibfnamefont {Yanning}\ \bibnamefont {Y.}},\ }\bibfield  {title} {\enquote {\bibinfo {title} {{Benchmarking of the Fock-space coupled-cluster method and uncertainty estimation: Magnetic hyperfine interaction in the excited state of BaF}},}\ }\href@noop {} {\bibfield  {journal} {\bibinfo  {journal} {Physical Review A}\ }\textbf {\bibinfo {volume} {105}},\ \bibinfo {pages} {052811} (\bibinfo {year} {2022})}\BibitemShut {NoStop}%
\bibitem [{\citenamefont {Kyuberis}\ \emph {et~al.}(2024)\citenamefont {Kyuberis}, \citenamefont {Pasteka}, \citenamefont {Eliav}, \citenamefont {Perrett}, \citenamefont {Sunaga}, \citenamefont {Udrescu}, \citenamefont {Wilkins}, \citenamefont {Ruiz},\ and\ \citenamefont {Borschevsky}}]{kyuberis2024theoretical}%
  \BibitemOpen
  \bibfield  {author} {\bibinfo {author} {\bibfnamefont {A.~A.}\ \bibnamefont {Kyuberis}}, \bibinfo {author} {\bibfnamefont {L.~F.}\ \bibnamefont {Pasteka}}, \bibinfo {author} {\bibfnamefont {E.}~\bibnamefont {Eliav}}, \bibinfo {author} {\bibfnamefont {H.~A.}\ \bibnamefont {Perrett}}, \bibinfo {author} {\bibfnamefont {A.}~\bibnamefont {Sunaga}}, \bibinfo {author} {\bibfnamefont {S.~M.}\ \bibnamefont {Udrescu}}, \bibinfo {author} {\bibfnamefont {S.~G.}\ \bibnamefont {Wilkins}}, \bibinfo {author} {\bibfnamefont {R.~F.~G.}\ \bibnamefont {Ruiz}}, \ and\ \bibinfo {author} {\bibfnamefont {A.}~\bibnamefont {Borschevsky}},\ }\bibfield  {title} {\enquote {\bibinfo {title} {{Theoretical determination of the ionization potentials of CaF, SrF, and BaF}},}\ }\href@noop {} {\bibfield  {journal} {\bibinfo  {journal} {Physical Review A}\ }\textbf {\bibinfo {volume} {109}},\ \bibinfo {pages} {022813} (\bibinfo {year} {2024})}\BibitemShut {NoStop}%
\bibitem [{\citenamefont {Chen}\ \emph {et~al.}(2016)\citenamefont {Chen}, \citenamefont {Bu},\ and\ \citenamefont {Yan}}]{Chen2016Structure}%
  \BibitemOpen
  \bibfield  {author} {\bibinfo {author} {\bibfnamefont {T.}~\bibnamefont {Chen}}, \bibinfo {author} {\bibfnamefont {W.}~\bibnamefont {Bu}}, \ and\ \bibinfo {author} {\bibfnamefont {B.}~\bibnamefont {Yan}},\ }\bibfield  {title} {\enquote {\bibinfo {title} {{Structure, branching ratios, and a laser-cooling scheme for the $^{138}$BaF molecule}},}\ }\href@noop {} {\bibfield  {journal} {\bibinfo  {journal} {Physical Review A}\ }\textbf {\bibinfo {volume} {94}},\ \bibinfo {pages} {063415} (\bibinfo {year} {2016})}\BibitemShut {NoStop}%
\bibitem [{\citenamefont {Chen}\ \emph {et~al.}(2019)\citenamefont {Chen}, \citenamefont {Bu},\ and\ \citenamefont {Yan}}]{chen2016erratum}%
  \BibitemOpen
  \bibfield  {author} {\bibinfo {author} {\bibfnamefont {T.}~\bibnamefont {Chen}}, \bibinfo {author} {\bibfnamefont {W.}~\bibnamefont {Bu}}, \ and\ \bibinfo {author} {\bibfnamefont {B.}~\bibnamefont {Yan}},\ }\bibfield  {title} {\enquote {\bibinfo {title} {{Erratum: Structure, branching ratios, and a laser-cooling scheme for the BaF 138 molecule [Phys. Rev. A 94, 063415 (2016)]}},}\ }\href@noop {} {\bibfield  {journal} {\bibinfo  {journal} {Physical Review A}\ }\textbf {\bibinfo {volume} {100}},\ \bibinfo {pages} {029901} (\bibinfo {year} {2019})}\BibitemShut {NoStop}%
\bibitem [{\citenamefont {Kang}\ \emph {et~al.}(2016)\citenamefont {Kang}, \citenamefont {Kuang}, \citenamefont {Jiang},\ and\ \citenamefont {Du}}]{Kang2016Suitability}%
  \BibitemOpen
  \bibfield  {author} {\bibinfo {author} {\bibfnamefont {S.}~\bibnamefont {Kang}}, \bibinfo {author} {\bibfnamefont {F.}~\bibnamefont {Kuang}}, \bibinfo {author} {\bibfnamefont {G.}~\bibnamefont {Jiang}}, \ and\ \bibinfo {author} {\bibfnamefont {J.}~\bibnamefont {Du}},\ }\bibfield  {title} {\enquote {\bibinfo {title} {{The suitability of barium monofluoride for laser cooling from ab initio study}},}\ }\href@noop {} {\bibfield  {journal} {\bibinfo  {journal} {Molecular Physics}\ }\textbf {\bibinfo {volume} {114}},\ \bibinfo {pages} {810} (\bibinfo {year} {2016})}\BibitemShut {NoStop}%
\bibitem [{\citenamefont {Xu}\ \emph {et~al.}(2017)\citenamefont {Xu}, \citenamefont {Wei}, \citenamefont {Xia}, \citenamefont {Deng},\ and\ \citenamefont {Yin}}]{xu2017baf}%
  \BibitemOpen
  \bibfield  {author} {\bibinfo {author} {\bibfnamefont {L.}~\bibnamefont {Xu}}, \bibinfo {author} {\bibfnamefont {B.}~\bibnamefont {Wei}}, \bibinfo {author} {\bibfnamefont {Y.}~\bibnamefont {Xia}}, \bibinfo {author} {\bibfnamefont {L.-Z.}\ \bibnamefont {Deng}}, \ and\ \bibinfo {author} {\bibfnamefont {J.-P.}\ \bibnamefont {Yin}},\ }\bibfield  {title} {\enquote {\bibinfo {title} {{BaF radical: A promising candidate for laser cooling and magneto-optical trapping}},}\ }\href@noop {} {\bibfield  {journal} {\bibinfo  {journal} {Chinese Physics B}\ }\textbf {\bibinfo {volume} {26}},\ \bibinfo {pages} {033702} (\bibinfo {year} {2017})}\BibitemShut {NoStop}%
\bibitem [{\citenamefont {Chen}\ \emph {et~al.}(2017)\citenamefont {Chen}, \citenamefont {Bu},\ and\ \citenamefont {Yan}}]{chen2017radiative}%
  \BibitemOpen
  \bibfield  {author} {\bibinfo {author} {\bibfnamefont {T.}~\bibnamefont {Chen}}, \bibinfo {author} {\bibfnamefont {W.}~\bibnamefont {Bu}}, \ and\ \bibinfo {author} {\bibfnamefont {B.}~\bibnamefont {Yan}},\ }\bibfield  {title} {\enquote {\bibinfo {title} {{Radiative deflection of a BaF molecular beam via optical cycling}},}\ }\href@noop {} {\bibfield  {journal} {\bibinfo  {journal} {Physical Review A}\ }\textbf {\bibinfo {volume} {96}},\ \bibinfo {pages} {053401} (\bibinfo {year} {2017})}\BibitemShut {NoStop}%
\bibitem [{\citenamefont {Hao}\ \emph {et~al.}(2019)\citenamefont {Hao}, \citenamefont {Pasteka}, \citenamefont {Visscher}, \citenamefont {Aggarwal}, \citenamefont {Bethlem}, \citenamefont {Boeschoten}, \citenamefont {Borschevsky}, \citenamefont {Denis}, \citenamefont {Esajas}, \citenamefont {Hoekstra}, \citenamefont {Jungmann}, \citenamefont {Marshall}, \citenamefont {Meijknecht}, \citenamefont {Mooij}, \citenamefont {Timmermans}, \citenamefont {Touwen}, \citenamefont {Ubachs}, \citenamefont {Willmann}, \citenamefont {Yin},\ and\ \citenamefont {Zapara1}}]{hao2019high}%
  \BibitemOpen
  \bibfield  {author} {\bibinfo {author} {\bibfnamefont {Y.}~\bibnamefont {Hao}}, \bibinfo {author} {\bibfnamefont {L.~F.}\ \bibnamefont {Pasteka}}, \bibinfo {author} {\bibfnamefont {L.}~\bibnamefont {Visscher}}, \bibinfo {author} {\bibfnamefont {P.}~\bibnamefont {Aggarwal}}, \bibinfo {author} {\bibfnamefont {H.~L.}\ \bibnamefont {Bethlem}}, \bibinfo {author} {\bibfnamefont {A.}~\bibnamefont {Boeschoten}}, \bibinfo {author} {\bibfnamefont {A.}~\bibnamefont {Borschevsky}}, \bibinfo {author} {\bibfnamefont {M.}~\bibnamefont {Denis}}, \bibinfo {author} {\bibfnamefont {K.}~\bibnamefont {Esajas}}, \bibinfo {author} {\bibfnamefont {S.}~\bibnamefont {Hoekstra}}, \bibinfo {author} {\bibfnamefont {K.}~\bibnamefont {Jungmann}}, \bibinfo {author} {\bibfnamefont {V.~R.}\ \bibnamefont {Marshall}}, \bibinfo {author} {\bibfnamefont {T.~B.}\ \bibnamefont {Meijknecht}}, \bibinfo {author} {\bibfnamefont {M.~C.}\ \bibnamefont {Mooij}}, \bibinfo {author} {\bibfnamefont {R.~G.~E.}\ \bibnamefont {Timmermans}}, \bibinfo {author}
  {\bibfnamefont {A.}~\bibnamefont {Touwen}}, \bibinfo {author} {\bibfnamefont {W.}~\bibnamefont {Ubachs}}, \bibinfo {author} {\bibfnamefont {L.}~\bibnamefont {Willmann}}, \bibinfo {author} {\bibfnamefont {Y.}~\bibnamefont {Yin}}, \ and\ \bibinfo {author} {\bibfnamefont {A.}~\bibnamefont {Zapara1}},\ }\bibfield  {title} {\enquote {\bibinfo {title} {{High accuracy theoretical investigations of CaF, SrF, and BaF and implications for laser-cooling}},}\ }\href@noop {} {\bibfield  {journal} {\bibinfo  {journal} {The Journal of Chemical Physics}\ }\textbf {\bibinfo {volume} {151}},\ \bibinfo {pages} {034302} (\bibinfo {year} {2019})}\BibitemShut {NoStop}%
\bibitem [{\citenamefont {Liang}\ \emph {et~al.}(2019)\citenamefont {Liang}, \citenamefont {Bu}, \citenamefont {Zhang}, \citenamefont {Chen},\ and\ \citenamefont {Yan}}]{liang2019improvements}%
  \BibitemOpen
  \bibfield  {author} {\bibinfo {author} {\bibfnamefont {Q.}~\bibnamefont {Liang}}, \bibinfo {author} {\bibfnamefont {W.}~\bibnamefont {Bu}}, \bibinfo {author} {\bibfnamefont {Y.}~\bibnamefont {Zhang}}, \bibinfo {author} {\bibfnamefont {T.}~\bibnamefont {Chen}}, \ and\ \bibinfo {author} {\bibfnamefont {B.}~\bibnamefont {Yan}},\ }\bibfield  {title} {\enquote {\bibinfo {title} {{Improvements on type-II Zeeman slowing of molecules through polarization selectivity}},}\ }\href@noop {} {\bibfield  {journal} {\bibinfo  {journal} {Physical Review A}\ }\textbf {\bibinfo {volume} {100}},\ \bibinfo {pages} {053402} (\bibinfo {year} {2019})}\BibitemShut {NoStop}%
\bibitem [{\citenamefont {Albrecht}\ \emph {et~al.}(2020)\citenamefont {Albrecht}, \citenamefont {Scharwaechter}, \citenamefont {Sixt}, \citenamefont {Hofer},\ and\ \citenamefont {Langen}}]{albrecht2020buffer}%
  \BibitemOpen
  \bibfield  {author} {\bibinfo {author} {\bibfnamefont {R.}~\bibnamefont {Albrecht}}, \bibinfo {author} {\bibfnamefont {M.}~\bibnamefont {Scharwaechter}}, \bibinfo {author} {\bibfnamefont {T.}~\bibnamefont {Sixt}}, \bibinfo {author} {\bibfnamefont {L.}~\bibnamefont {Hofer}}, \ and\ \bibinfo {author} {\bibfnamefont {T.}~\bibnamefont {Langen}},\ }\bibfield  {title} {\enquote {\bibinfo {title} {{Buffer-gas cooling, high-resolution spectroscopy, and optical cycling of barium monofluoride molecules}},}\ }\href@noop {} {\bibfield  {journal} {\bibinfo  {journal} {Physical Review A}\ }\textbf {\bibinfo {volume} {101}},\ \bibinfo {pages} {013413} (\bibinfo {year} {2020})}\BibitemShut {NoStop}%
\bibitem [{\citenamefont {Yang}\ \emph {et~al.}(2020)\citenamefont {Yang}, \citenamefont {Tang},\ and\ \citenamefont {Han}}]{yang2020ab}%
  \BibitemOpen
  \bibfield  {author} {\bibinfo {author} {\bibfnamefont {R.}~\bibnamefont {Yang}}, \bibinfo {author} {\bibfnamefont {B.}~\bibnamefont {Tang}}, \ and\ \bibinfo {author} {\bibfnamefont {X.}~\bibnamefont {Han}},\ }\bibfield  {title} {\enquote {\bibinfo {title} {{Ab initio theory study of laser cooling of barium monohalides}},}\ }\href@noop {} {\bibfield  {journal} {\bibinfo  {journal} {Royal Society of Chemistry Advances}\ }\textbf {\bibinfo {volume} {10}},\ \bibinfo {pages} {20778} (\bibinfo {year} {2020})}\BibitemShut {NoStop}%
\bibitem [{\citenamefont {Kogel}\ \emph {et~al.}(2021)\citenamefont {Kogel}, \citenamefont {Rockenh{\"a}user}, \citenamefont {Albrecht},\ and\ \citenamefont {Langen}}]{kogel2021laser}%
  \BibitemOpen
  \bibfield  {author} {\bibinfo {author} {\bibfnamefont {F.}~\bibnamefont {Kogel}}, \bibinfo {author} {\bibfnamefont {M.}~\bibnamefont {Rockenh{\"a}user}}, \bibinfo {author} {\bibfnamefont {R.}~\bibnamefont {Albrecht}}, \ and\ \bibinfo {author} {\bibfnamefont {T.}~\bibnamefont {Langen}},\ }\bibfield  {title} {\enquote {\bibinfo {title} {{A laser cooling scheme for precision measurements using fermionic barium monofluoride ($^{137}$Ba$^{19}$F) molecules}},}\ }\href@noop {} {\bibfield  {journal} {\bibinfo  {journal} {New Journal of Physics}\ }\textbf {\bibinfo {volume} {23}},\ \bibinfo {pages} {095003} (\bibinfo {year} {2021})}\BibitemShut {NoStop}%
\bibitem [{\citenamefont {Zhang}\ \emph {et~al.}(2022)\citenamefont {Zhang}, \citenamefont {Zeng}, \citenamefont {Liang}, \citenamefont {Bu},\ and\ \citenamefont {Yan}}]{zhang2022doppler}%
  \BibitemOpen
  \bibfield  {author} {\bibinfo {author} {\bibfnamefont {Y.}~\bibnamefont {Zhang}}, \bibinfo {author} {\bibfnamefont {Z.}~\bibnamefont {Zeng}}, \bibinfo {author} {\bibfnamefont {Q.}~\bibnamefont {Liang}}, \bibinfo {author} {\bibfnamefont {W.}~\bibnamefont {Bu}}, \ and\ \bibinfo {author} {\bibfnamefont {B.}~\bibnamefont {Yan}},\ }\bibfield  {title} {\enquote {\bibinfo {title} {{Doppler cooling of buffer-gas-cooled barium monofluoride molecules}},}\ }\href@noop {} {\bibfield  {journal} {\bibinfo  {journal} {Physical Review A}\ }\textbf {\bibinfo {volume} {105}},\ \bibinfo {pages} {033307} (\bibinfo {year} {2022})}\BibitemShut {NoStop}%
\bibitem [{\citenamefont {Marsman}\ \emph {et~al.}(2023{\natexlab{a}})\citenamefont {Marsman}, \citenamefont {Heinrich}, \citenamefont {Horbatsch},\ and\ \citenamefont {Hessels}}]{marsman2023large}%
  \BibitemOpen
  \bibfield  {author} {\bibinfo {author} {\bibfnamefont {A.}~\bibnamefont {Marsman}}, \bibinfo {author} {\bibfnamefont {D.}~\bibnamefont {Heinrich}}, \bibinfo {author} {\bibfnamefont {M.}~\bibnamefont {Horbatsch}}, \ and\ \bibinfo {author} {\bibfnamefont {E.~A.}\ \bibnamefont {Hessels}},\ }\bibfield  {title} {\enquote {\bibinfo {title} {{Large optical forces on a barium monofluoride molecule using laser pulses for absorption and stimulated emission: A full density-matrix simulation}},}\ }\href@noop {} {\bibfield  {journal} {\bibinfo  {journal} {Physical Review A}\ }\textbf {\bibinfo {volume} {107}},\ \bibinfo {pages} {032811} (\bibinfo {year} {2023}{\natexlab{a}})}\BibitemShut {NoStop}%
\bibitem [{\citenamefont {Marsman}\ \emph {et~al.}(2023{\natexlab{b}})\citenamefont {Marsman}, \citenamefont {Horbatsch},\ and\ \citenamefont {Hessels}}]{marsman2023deflection}%
  \BibitemOpen
  \bibfield  {author} {\bibinfo {author} {\bibfnamefont {A.}~\bibnamefont {Marsman}}, \bibinfo {author} {\bibfnamefont {M.}~\bibnamefont {Horbatsch}}, \ and\ \bibinfo {author} {\bibfnamefont {E.~A.}\ \bibnamefont {Hessels}},\ }\bibfield  {title} {\enquote {\bibinfo {title} {{Deflection of barium monofluoride molecules using the bichromatic force: A density-matrix simulation}},}\ }\href@noop {} {\bibfield  {journal} {\bibinfo  {journal} {Physical Review A}\ }\textbf {\bibinfo {volume} {108}},\ \bibinfo {pages} {012811} (\bibinfo {year} {2023}{\natexlab{b}})}\BibitemShut {NoStop}%
\bibitem [{\citenamefont {Rockenh{\"a}user}\ \emph {et~al.}(2023)\citenamefont {Rockenh{\"a}user}, \citenamefont {Kogel}, \citenamefont {Pultinevicius},\ and\ \citenamefont {Langen}}]{rockenhauser2023absorption}%
  \BibitemOpen
  \bibfield  {author} {\bibinfo {author} {\bibfnamefont {M.}~\bibnamefont {Rockenh{\"a}user}}, \bibinfo {author} {\bibfnamefont {F.}~\bibnamefont {Kogel}}, \bibinfo {author} {\bibfnamefont {E.}~\bibnamefont {Pultinevicius}}, \ and\ \bibinfo {author} {\bibfnamefont {T.}~\bibnamefont {Langen}},\ }\bibfield  {title} {\enquote {\bibinfo {title} {{Absorption spectroscopy for laser cooling and high-fidelity detection of barium monofluoride molecules}},}\ }\href@noop {} {\bibfield  {journal} {\bibinfo  {journal} {Physical Review A}\ }\textbf {\bibinfo {volume} {108}},\ \bibinfo {pages} {062812} (\bibinfo {year} {2023})}\BibitemShut {NoStop}%
\bibitem [{\citenamefont {Rockenh{\"a}user}\ \emph {et~al.}(2024)\citenamefont {Rockenh{\"a}user}, \citenamefont {Kogel}, \citenamefont {Garg}, \citenamefont {Morales-Ram{\'\i}rez},\ and\ \citenamefont {Langen}}]{rockenhauser2024laser}%
  \BibitemOpen
  \bibfield  {author} {\bibinfo {author} {\bibfnamefont {M.}~\bibnamefont {Rockenh{\"a}user}}, \bibinfo {author} {\bibfnamefont {F.}~\bibnamefont {Kogel}}, \bibinfo {author} {\bibfnamefont {T.}~\bibnamefont {Garg}}, \bibinfo {author} {\bibfnamefont {S.~A.}\ \bibnamefont {Morales-Ram{\'\i}rez}}, \ and\ \bibinfo {author} {\bibfnamefont {T.}~\bibnamefont {Langen}},\ }\bibfield  {title} {\enquote {\bibinfo {title} {{Laser cooling of barium monofluoride molecules using synthesized optical spectra}},}\ }\href@noop {} {\bibfield  {journal} {\bibinfo  {journal} {arXiv preprint arXiv:2405.09427}\ } (\bibinfo {year} {2024})}\BibitemShut {NoStop}%
\bibitem [{\citenamefont {Zeng}\ \emph {et~al.}(2024)\citenamefont {Zeng}, \citenamefont {Deng}, \citenamefont {Yang},\ and\ \citenamefont {Yan}}]{zeng2024three}%
  \BibitemOpen
  \bibfield  {author} {\bibinfo {author} {\bibfnamefont {Z.}~\bibnamefont {Zeng}}, \bibinfo {author} {\bibfnamefont {S.}~\bibnamefont {Deng}}, \bibinfo {author} {\bibfnamefont {S.}~\bibnamefont {Yang}}, \ and\ \bibinfo {author} {\bibfnamefont {B.}~\bibnamefont {Yan}},\ }\bibfield  {title} {\enquote {\bibinfo {title} {Three-dimensional magneto-optical trapping of barium monofluoride},}\ }\href@noop {} {\bibfield  {journal} {\bibinfo  {journal} {arXiv preprint arXiv:2405.17883}\ } (\bibinfo {year} {2024})}\BibitemShut {NoStop}%
\bibitem [{\citenamefont {Kogel}\ \emph {et~al.}(2024)\citenamefont {Kogel}, \citenamefont {Garg}, \citenamefont {Rockenh{\"a}user}, \citenamefont {Morales-Ram{\'\i}rez},\ and\ \citenamefont {Langen}}]{kogel2024isotopologue}%
  \BibitemOpen
  \bibfield  {author} {\bibinfo {author} {\bibfnamefont {F.}~\bibnamefont {Kogel}}, \bibinfo {author} {\bibfnamefont {T.}~\bibnamefont {Garg}}, \bibinfo {author} {\bibfnamefont {M.}~\bibnamefont {Rockenh{\"a}user}}, \bibinfo {author} {\bibfnamefont {S.~A.}\ \bibnamefont {Morales-Ram{\'\i}rez}}, \ and\ \bibinfo {author} {\bibfnamefont {T.}~\bibnamefont {Langen}},\ }\bibfield  {title} {\enquote {\bibinfo {title} {{Isotopologue-selective laser cooling of molecules}},}\ }\href@noop {} {\bibfield  {journal} {\bibinfo  {journal} {arXiv preprint arXiv:2406.01569}\ } (\bibinfo {year} {2024})}\BibitemShut {NoStop}%
\bibitem [{\citenamefont {Kozlov}(1985)}]{kozlov1985semiempirical}%
  \BibitemOpen
  \bibfield  {author} {\bibinfo {author} {\bibfnamefont {M.~G.}\ \bibnamefont {Kozlov}},\ }\bibfield  {title} {\enquote {\bibinfo {title} {{Semiempirical calculations of P- and P,$\Gamma$-odd effects in diatomic molecules-radicals}},}\ }\href@noop {} {\bibfield  {journal} {\bibinfo  {journal} {Zh. Eksp. Teor. Fiz}\ }\textbf {\bibinfo {volume} {89}},\ \bibinfo {pages} {1933} (\bibinfo {year} {1985})}\BibitemShut {NoStop}%
\bibitem [{\citenamefont {Kozlov}\ \emph {et~al.}(1997)\citenamefont {Kozlov}, \citenamefont {Titov}, \citenamefont {Mosyagin},\ and\ \citenamefont {Souchko}}]{kozlov1997enhancement}%
  \BibitemOpen
  \bibfield  {author} {\bibinfo {author} {\bibfnamefont {M.~G.}\ \bibnamefont {Kozlov}}, \bibinfo {author} {\bibfnamefont {A.~V.}\ \bibnamefont {Titov}}, \bibinfo {author} {\bibfnamefont {N.~S.}\ \bibnamefont {Mosyagin}}, \ and\ \bibinfo {author} {\bibfnamefont {P.~V.}\ \bibnamefont {Souchko}},\ }\bibfield  {title} {\enquote {\bibinfo {title} {{Enhancement of the electric dipole moment of the electron in the BaF molecule}},}\ }\href@noop {} {\bibfield  {journal} {\bibinfo  {journal} {Physical Review A}\ }\textbf {\bibinfo {volume} {56}},\ \bibinfo {pages} {R3326} (\bibinfo {year} {1997})}\BibitemShut {NoStop}%
\bibitem [{\citenamefont {Meyer}\ \emph {et~al.}(2006)\citenamefont {Meyer}, \citenamefont {Bohn},\ and\ \citenamefont {Deskevich}}]{meyer2006candidate}%
  \BibitemOpen
  \bibfield  {author} {\bibinfo {author} {\bibfnamefont {E.~R.}\ \bibnamefont {Meyer}}, \bibinfo {author} {\bibfnamefont {J.~L.}\ \bibnamefont {Bohn}}, \ and\ \bibinfo {author} {\bibfnamefont {M.~P.}\ \bibnamefont {Deskevich}},\ }\bibfield  {title} {\enquote {\bibinfo {title} {{Candidate molecular ions for an electron electric dipole moment experiment}},}\ }\href@noop {} {\bibfield  {journal} {\bibinfo  {journal} {Physical Review A}\ }\textbf {\bibinfo {volume} {73}},\ \bibinfo {pages} {062108} (\bibinfo {year} {2006})}\BibitemShut {NoStop}%
\bibitem [{\citenamefont {Nayak}\ and\ \citenamefont {Chaudhuri}(2006)}]{nayak2006ab}%
  \BibitemOpen
  \bibfield  {author} {\bibinfo {author} {\bibfnamefont {M.~K.}\ \bibnamefont {Nayak}}\ and\ \bibinfo {author} {\bibfnamefont {R.~K.}\ \bibnamefont {Chaudhuri}},\ }\bibfield  {title} {\enquote {\bibinfo {title} {{Ab initio calculation of P, T-odd interaction constant in BaF: a restricted active space configuration interaction approach}},}\ }\href@noop {} {\bibfield  {journal} {\bibinfo  {journal} {Journal of Physics B: Atomic, Molecular and Optical Physics}\ }\textbf {\bibinfo {volume} {39}},\ \bibinfo {pages} {1231} (\bibinfo {year} {2006})}\BibitemShut {NoStop}%
\bibitem [{\citenamefont {Nayak}\ and\ \citenamefont {Chaudhuri}(2008)}]{nayak2008calculation}%
  \BibitemOpen
  \bibfield  {author} {\bibinfo {author} {\bibfnamefont {M.~K.}\ \bibnamefont {Nayak}}\ and\ \bibinfo {author} {\bibfnamefont {R.~K.}\ \bibnamefont {Chaudhuri}},\ }\bibfield  {title} {\enquote {\bibinfo {title} {{Calculation of the electron-nucleus scalar-pseudoscalar interaction constant $W_S$ for YbF and BaF molecules: A perturbative approach}},}\ }\href@noop {} {\bibfield  {journal} {\bibinfo  {journal} {Physical Review A}\ }\textbf {\bibinfo {volume} {78}},\ \bibinfo {pages} {012506} (\bibinfo {year} {2008})}\BibitemShut {NoStop}%
\bibitem [{\citenamefont {Nayak}\ and\ \citenamefont {Das}(2009)}]{nayak2009relativistic}%
  \BibitemOpen
  \bibfield  {author} {\bibinfo {author} {\bibfnamefont {M.~K.}\ \bibnamefont {Nayak}}\ and\ \bibinfo {author} {\bibfnamefont {B.~P.}\ \bibnamefont {Das}},\ }\bibfield  {title} {\enquote {\bibinfo {title} {{Relativistic configuration-interaction study of the nuclear-spin-dependent parity-nonconserving electron-nucleus interaction constant $W_A$ in BaF}},}\ }\href@noop {} {\bibfield  {journal} {\bibinfo  {journal} {Physical Review A}\ }\textbf {\bibinfo {volume} {79}},\ \bibinfo {pages} {060502} (\bibinfo {year} {2009})}\BibitemShut {NoStop}%
\bibitem [{\citenamefont {Flambaum}\ \emph {et~al.}(2014)\citenamefont {Flambaum}, \citenamefont {DeMille},\ and\ \citenamefont {Kozlov}}]{flambaum2014time}%
  \BibitemOpen
  \bibfield  {author} {\bibinfo {author} {\bibfnamefont {V.~V.}\ \bibnamefont {Flambaum}}, \bibinfo {author} {\bibfnamefont {D.}~\bibnamefont {DeMille}}, \ and\ \bibinfo {author} {\bibfnamefont {M.~G.}\ \bibnamefont {Kozlov}},\ }\bibfield  {title} {\enquote {\bibinfo {title} {{Time-reversal symmetry violation in molecules induced by nuclear magnetic quadrupole moments}},}\ }\href@noop {} {\bibfield  {journal} {\bibinfo  {journal} {Physical Review Letters}\ }\textbf {\bibinfo {volume} {113}},\ \bibinfo {pages} {103003} (\bibinfo {year} {2014})}\BibitemShut {NoStop}%
\bibitem [{\citenamefont {Fukuda}\ \emph {et~al.}(2016)\citenamefont {Fukuda}, \citenamefont {Soga}, \citenamefont {Senami},\ and\ \citenamefont {Tachibana}}]{fukuda2016local}%
  \BibitemOpen
  \bibfield  {author} {\bibinfo {author} {\bibfnamefont {M.}~\bibnamefont {Fukuda}}, \bibinfo {author} {\bibfnamefont {K.}~\bibnamefont {Soga}}, \bibinfo {author} {\bibfnamefont {M.}~\bibnamefont {Senami}}, \ and\ \bibinfo {author} {\bibfnamefont {A.}~\bibnamefont {Tachibana}},\ }\bibfield  {title} {\enquote {\bibinfo {title} {{Local spin dynamics with the electron electric dipole moment}},}\ }\href@noop {} {\bibfield  {journal} {\bibinfo  {journal} {Physical Review A}\ }\textbf {\bibinfo {volume} {93}},\ \bibinfo {pages} {012518} (\bibinfo {year} {2016})}\BibitemShut {NoStop}%
\bibitem [{\citenamefont {Hao}\ \emph {et~al.}(2018)\citenamefont {Hao}, \citenamefont {Ilia\v{s}}, \citenamefont {Eliav}, \citenamefont {Schwerdtfeger}, \citenamefont {Flambaum},\ and\ \citenamefont {Borschevsky}}]{hao2018nuclear}%
  \BibitemOpen
  \bibfield  {author} {\bibinfo {author} {\bibfnamefont {Y.}~\bibnamefont {Hao}}, \bibinfo {author} {\bibfnamefont {M.}~\bibnamefont {Ilia\v{s}}}, \bibinfo {author} {\bibfnamefont {E.}~\bibnamefont {Eliav}}, \bibinfo {author} {\bibfnamefont {P.}~\bibnamefont {Schwerdtfeger}}, \bibinfo {author} {\bibfnamefont {V.~V.}\ \bibnamefont {Flambaum}}, \ and\ \bibinfo {author} {\bibfnamefont {A.}~\bibnamefont {Borschevsky}},\ }\bibfield  {title} {\enquote {\bibinfo {title} {{Nuclear anapole moment interaction in BaF from relativistic coupled-cluster theory}},}\ }\href@noop {} {\bibfield  {journal} {\bibinfo  {journal} {Physical Review A}\ }\textbf {\bibinfo {volume} {98}},\ \bibinfo {pages} {032510} (\bibinfo {year} {2018})}\BibitemShut {NoStop}%
\bibitem [{\citenamefont {Altunta{\c{s}}}\ \emph {et~al.}(2018)\citenamefont {Altunta{\c{s}}}, \citenamefont {Ammon}, \citenamefont {Cahn},\ and\ \citenamefont {DeMille}}]{altuntacs2018demonstration}%
  \BibitemOpen
  \bibfield  {author} {\bibinfo {author} {\bibfnamefont {E.}~\bibnamefont {Altunta{\c{s}}}}, \bibinfo {author} {\bibfnamefont {J.}~\bibnamefont {Ammon}}, \bibinfo {author} {\bibfnamefont {S.~B.}\ \bibnamefont {Cahn}}, \ and\ \bibinfo {author} {\bibfnamefont {D.}~\bibnamefont {DeMille}},\ }\bibfield  {title} {\enquote {\bibinfo {title} {{Demonstration of a sensitive method to measure nuclear-spin-dependent parity violation}},}\ }\href@noop {} {\bibfield  {journal} {\bibinfo  {journal} {Physical Review Letters}\ }\textbf {\bibinfo {volume} {120}},\ \bibinfo {pages} {142501} (\bibinfo {year} {2018})}\BibitemShut {NoStop}%
\bibitem [{\citenamefont {Abe}\ \emph {et~al.}(2018)\citenamefont {Abe}, \citenamefont {Prasannaa},\ and\ \citenamefont {Das}}]{abe2018application}%
  \BibitemOpen
  \bibfield  {author} {\bibinfo {author} {\bibfnamefont {M.}~\bibnamefont {Abe}}, \bibinfo {author} {\bibfnamefont {V.~S.}\ \bibnamefont {Prasannaa}}, \ and\ \bibinfo {author} {\bibfnamefont {B.~P.}\ \bibnamefont {Das}},\ }\bibfield  {title} {\enquote {\bibinfo {title} {{Application of the finite-field coupled-cluster method to calculate molecular properties relevant to electron electric-dipole-moment searches}},}\ }\href@noop {} {\bibfield  {journal} {\bibinfo  {journal} {Physical Review A}\ }\textbf {\bibinfo {volume} {97}},\ \bibinfo {pages} {032515} (\bibinfo {year} {2018})}\BibitemShut {NoStop}%
\bibitem [{\citenamefont {Vutha}\ \emph {et~al.}(2018{\natexlab{a}})\citenamefont {Vutha}, \citenamefont {Horbatsch},\ and\ \citenamefont {Hessels}}]{vutha2018oriented}%
  \BibitemOpen
  \bibfield  {author} {\bibinfo {author} {\bibfnamefont {A.~C.}\ \bibnamefont {Vutha}}, \bibinfo {author} {\bibfnamefont {M.}~\bibnamefont {Horbatsch}}, \ and\ \bibinfo {author} {\bibfnamefont {E.~A.}\ \bibnamefont {Hessels}},\ }\bibfield  {title} {\enquote {\bibinfo {title} {{Oriented polar molecules in a solid inert-gas matrix: a proposed method for measuring the electric dipole moment of the electron}},}\ }\href@noop {} {\bibfield  {journal} {\bibinfo  {journal} {Atoms}\ }\textbf {\bibinfo {volume} {6}},\ \bibinfo {pages} {3} (\bibinfo {year} {2018}{\natexlab{a}})}\BibitemShut {NoStop}%
\bibitem [{\citenamefont {Vutha}\ \emph {et~al.}(2018{\natexlab{b}})\citenamefont {Vutha}, \citenamefont {Horbatsch},\ and\ \citenamefont {Hessels}}]{vutha2018orientation}%
  \BibitemOpen
  \bibfield  {author} {\bibinfo {author} {\bibfnamefont {A.~C.}\ \bibnamefont {Vutha}}, \bibinfo {author} {\bibfnamefont {M.}~\bibnamefont {Horbatsch}}, \ and\ \bibinfo {author} {\bibfnamefont {E.~A.}\ \bibnamefont {Hessels}},\ }\bibfield  {title} {\enquote {\bibinfo {title} {{Orientation-dependent hyperfine structure of polar molecules in a rare-gas matrix: A scheme for measuring the electron electric dipole moment}},}\ }\href@noop {} {\bibfield  {journal} {\bibinfo  {journal} {Physical Review A}\ }\textbf {\bibinfo {volume} {98}},\ \bibinfo {pages} {032513} (\bibinfo {year} {2018}{\natexlab{b}})}\BibitemShut {NoStop}%
\bibitem [{\citenamefont {Aggarwal}\ \emph {et~al.}(2018)\citenamefont {Aggarwal}, \citenamefont {Bethlem}, \citenamefont {Borschevsky}, \citenamefont {Denis}, \citenamefont {Esajas}, \citenamefont {Haase}, \citenamefont {Hao}, \citenamefont {Hoekstra}, \citenamefont {Jungmann}, \citenamefont {Meijknecht}, \citenamefont {Mooij}, \citenamefont {Timmermans}, \citenamefont {Ubachs}, \citenamefont {Willmann},\ and\ \citenamefont {Zapara}}]{aggarwal2018measuring}%
  \BibitemOpen
  \bibfield  {author} {\bibinfo {author} {\bibfnamefont {P.}~\bibnamefont {Aggarwal}}, \bibinfo {author} {\bibfnamefont {H.~L.}\ \bibnamefont {Bethlem}}, \bibinfo {author} {\bibfnamefont {A.}~\bibnamefont {Borschevsky}}, \bibinfo {author} {\bibfnamefont {M.}~\bibnamefont {Denis}}, \bibinfo {author} {\bibfnamefont {K.}~\bibnamefont {Esajas}}, \bibinfo {author} {\bibfnamefont {P.~A.~B.}\ \bibnamefont {Haase}}, \bibinfo {author} {\bibfnamefont {Y.}~\bibnamefont {Hao}}, \bibinfo {author} {\bibfnamefont {S.}~\bibnamefont {Hoekstra}}, \bibinfo {author} {\bibfnamefont {K.}~\bibnamefont {Jungmann}}, \bibinfo {author} {\bibfnamefont {T.~B.}\ \bibnamefont {Meijknecht}}, \bibinfo {author} {\bibfnamefont {M.~C.}\ \bibnamefont {Mooij}}, \bibinfo {author} {\bibfnamefont {R.~G.~E.}\ \bibnamefont {Timmermans}}, \bibinfo {author} {\bibfnamefont {W.}~\bibnamefont {Ubachs}}, \bibinfo {author} {\bibfnamefont {L.}~\bibnamefont {Willmann}}, \ and\ \bibinfo {author} {\bibfnamefont {A.}~\bibnamefont {Zapara}},\ }\bibfield  {title}
  {\enquote {\bibinfo {title} {{Measuring the electric dipole moment of the electron in BaF}},}\ }\href@noop {} {\bibfield  {journal} {\bibinfo  {journal} {The European Physical Journal D}\ }\textbf {\bibinfo {volume} {72}},\ \bibinfo {pages} {197} (\bibinfo {year} {2018})}\BibitemShut {NoStop}%
\bibitem [{\citenamefont {Prasannaa}\ \emph {et~al.}(2019)\citenamefont {Prasannaa}, \citenamefont {Sunaga}, \citenamefont {Abe}, \citenamefont {Hada}, \citenamefont {Shitara}, \citenamefont {Sakurai},\ and\ \citenamefont {Das}}]{prasannaa2019role}%
  \BibitemOpen
  \bibfield  {author} {\bibinfo {author} {\bibfnamefont {V.~S.}\ \bibnamefont {Prasannaa}}, \bibinfo {author} {\bibfnamefont {A.}~\bibnamefont {Sunaga}}, \bibinfo {author} {\bibfnamefont {M.}~\bibnamefont {Abe}}, \bibinfo {author} {\bibfnamefont {M.}~\bibnamefont {Hada}}, \bibinfo {author} {\bibfnamefont {N.}~\bibnamefont {Shitara}}, \bibinfo {author} {\bibfnamefont {A.}~\bibnamefont {Sakurai}}, \ and\ \bibinfo {author} {\bibfnamefont {B.~P.}\ \bibnamefont {Das}},\ }\bibfield  {title} {\enquote {\bibinfo {title} {{The role of relativistic many-body theory in electron electric dipole moment searches using cold molecules}},}\ }\href@noop {} {\bibfield  {journal} {\bibinfo  {journal} {Atoms}\ }\textbf {\bibinfo {volume} {7}},\ \bibinfo {pages} {58} (\bibinfo {year} {2019})}\BibitemShut {NoStop}%
\bibitem [{\citenamefont {Denis}\ \emph {et~al.}(2020)\citenamefont {Denis}, \citenamefont {Hao}, \citenamefont {Eliav}, \citenamefont {Hutzler}, \citenamefont {Nayak}, \citenamefont {Timmermans},\ and\ \citenamefont {Borschesvky}}]{denis2020enhanced}%
  \BibitemOpen
  \bibfield  {author} {\bibinfo {author} {\bibfnamefont {M.}~\bibnamefont {Denis}}, \bibinfo {author} {\bibfnamefont {Y.}~\bibnamefont {Hao}}, \bibinfo {author} {\bibfnamefont {E.}~\bibnamefont {Eliav}}, \bibinfo {author} {\bibfnamefont {N.~R.}\ \bibnamefont {Hutzler}}, \bibinfo {author} {\bibfnamefont {M.~K.}\ \bibnamefont {Nayak}}, \bibinfo {author} {\bibfnamefont {R.~G.~E.}\ \bibnamefont {Timmermans}}, \ and\ \bibinfo {author} {\bibfnamefont {A.}~\bibnamefont {Borschesvky}},\ }\bibfield  {title} {\enquote {\bibinfo {title} {{Enhanced P, T-violating nuclear magnetic quadrupole moment effects in laser-coolable molecules}},}\ }\href@noop {} {\bibfield  {journal} {\bibinfo  {journal} {The Journal of Chemical Physics}\ }\textbf {\bibinfo {volume} {152}} (\bibinfo {year} {2020})}\BibitemShut {NoStop}%
\bibitem [{\citenamefont {Talukdar}\ \emph {et~al.}(2020)\citenamefont {Talukdar}, \citenamefont {Nayak}, \citenamefont {Vaval},\ and\ \citenamefont {Pal}}]{talukdar2020relativistic}%
  \BibitemOpen
  \bibfield  {author} {\bibinfo {author} {\bibfnamefont {K.}~\bibnamefont {Talukdar}}, \bibinfo {author} {\bibfnamefont {M.~K.}\ \bibnamefont {Nayak}}, \bibinfo {author} {\bibfnamefont {N.}~\bibnamefont {Vaval}}, \ and\ \bibinfo {author} {\bibfnamefont {S.}~\bibnamefont {Pal}},\ }\bibfield  {title} {\enquote {\bibinfo {title} {{Relativistic coupled-cluster study of BaF in search of violation}},}\ }\href@noop {} {\bibfield  {journal} {\bibinfo  {journal} {Journal of Physics B: Atomic, Molecular and Optical Physics}\ }\textbf {\bibinfo {volume} {53}},\ \bibinfo {pages} {135102} (\bibinfo {year} {2020})}\BibitemShut {NoStop}%
\bibitem [{\citenamefont {Haase}\ \emph {et~al.}(2021)\citenamefont {Haase}, \citenamefont {Doeglas}, \citenamefont {Boeschoten}, \citenamefont {Eliav}, \citenamefont {Ilia{\v{s}}}, \citenamefont {Aggarwal}, \citenamefont {Bethlem}, \citenamefont {Borschevsky}, \citenamefont {Esajas}, \citenamefont {Hao}, \citenamefont {Hoekstra}, \citenamefont {Marshall}, \citenamefont {Meijknecht}, \citenamefont {Mooij}, \citenamefont {Steinebach}, \citenamefont {Timmermans}, \citenamefont {Touwen}, \citenamefont {Ubachs}, \citenamefont {Willmann},\ and\ \citenamefont {Yin}}]{haase2021systematic}%
  \BibitemOpen
  \bibfield  {author} {\bibinfo {author} {\bibfnamefont {P.~A.~B.}\ \bibnamefont {Haase}}, \bibinfo {author} {\bibfnamefont {D.~J.}\ \bibnamefont {Doeglas}}, \bibinfo {author} {\bibfnamefont {A.}~\bibnamefont {Boeschoten}}, \bibinfo {author} {\bibfnamefont {E.}~\bibnamefont {Eliav}}, \bibinfo {author} {\bibfnamefont {M.}~\bibnamefont {Ilia{\v{s}}}}, \bibinfo {author} {\bibfnamefont {P.}~\bibnamefont {Aggarwal}}, \bibinfo {author} {\bibfnamefont {H.~L.}\ \bibnamefont {Bethlem}}, \bibinfo {author} {\bibfnamefont {A.}~\bibnamefont {Borschevsky}}, \bibinfo {author} {\bibfnamefont {K.}~\bibnamefont {Esajas}}, \bibinfo {author} {\bibfnamefont {Y.}~\bibnamefont {Hao}}, \bibinfo {author} {\bibfnamefont {S.}~\bibnamefont {Hoekstra}}, \bibinfo {author} {\bibfnamefont {V.~R.}\ \bibnamefont {Marshall}}, \bibinfo {author} {\bibfnamefont {T.~B.}\ \bibnamefont {Meijknecht}}, \bibinfo {author} {\bibfnamefont {M.~C.}\ \bibnamefont {Mooij}}, \bibinfo {author} {\bibfnamefont {K.}~\bibnamefont {Steinebach}}, \bibinfo {author}
  {\bibfnamefont {;~R. G.~E.}\ \bibnamefont {Timmermans}}, \bibinfo {author} {\bibfnamefont {A.~P.}\ \bibnamefont {Touwen}}, \bibinfo {author} {\bibfnamefont {W.}~\bibnamefont {Ubachs}}, \bibinfo {author} {\bibfnamefont {L.}~\bibnamefont {Willmann}}, \ and\ \bibinfo {author} {\bibfnamefont {Y.}~\bibnamefont {Yin}},\ }\bibfield  {title} {\enquote {\bibinfo {title} {{Systematic study and uncertainty evaluation of P, T-odd molecular enhancement factors in BaF}},}\ }\href@noop {} {\bibfield  {journal} {\bibinfo  {journal} {The Journal of Chemical Physics}\ }\textbf {\bibinfo {volume} {155}} (\bibinfo {year} {2021})}\BibitemShut {NoStop}%
\bibitem [{\citenamefont {Prosnyak}\ and\ \citenamefont {Skripnikov}(2024)}]{prosnyak2024axion}%
  \BibitemOpen
  \bibfield  {author} {\bibinfo {author} {\bibfnamefont {S.~D.}\ \bibnamefont {Prosnyak}}\ and\ \bibinfo {author} {\bibfnamefont {L.~V.}\ \bibnamefont {Skripnikov}},\ }\bibfield  {title} {\enquote {\bibinfo {title} {{Axion-mediated electron-nucleus and electron-electron interactions in the barium monofluoride molecule}},}\ }\href@noop {} {\bibfield  {journal} {\bibinfo  {journal} {Physical Review A}\ }\textbf {\bibinfo {volume} {109}},\ \bibinfo {pages} {042821} (\bibinfo {year} {2024})}\BibitemShut {NoStop}%
\bibitem [{\citenamefont {Boeschoten}\ \emph {et~al.}(2024)\citenamefont {Boeschoten}, \citenamefont {Marshall}, \citenamefont {Meijknecht}, \citenamefont {Touwen}, \citenamefont {Bethlem}, \citenamefont {Borschevsky}, \citenamefont {Hoekstra}, \citenamefont {van Hofslot}, \citenamefont {Jungmann}, \citenamefont {Mooij}, \citenamefont {Timmermans}, \citenamefont {Ubachs},\ and\ \citenamefont {Willmann}}]{boeschoten2024spin}%
  \BibitemOpen
  \bibfield  {author} {\bibinfo {author} {\bibfnamefont {A.}~\bibnamefont {Boeschoten}}, \bibinfo {author} {\bibfnamefont {V.~R.}\ \bibnamefont {Marshall}}, \bibinfo {author} {\bibfnamefont {T.~B.}\ \bibnamefont {Meijknecht}}, \bibinfo {author} {\bibfnamefont {A.}~\bibnamefont {Touwen}}, \bibinfo {author} {\bibfnamefont {H.~L.}\ \bibnamefont {Bethlem}}, \bibinfo {author} {\bibfnamefont {A.}~\bibnamefont {Borschevsky}}, \bibinfo {author} {\bibfnamefont {S.}~\bibnamefont {Hoekstra}}, \bibinfo {author} {\bibfnamefont {J.~W.~F.}\ \bibnamefont {van Hofslot}}, \bibinfo {author} {\bibfnamefont {K.}~\bibnamefont {Jungmann}}, \bibinfo {author} {\bibfnamefont {M.~C.}\ \bibnamefont {Mooij}}, \bibinfo {author} {\bibfnamefont {R.~G.~E.}\ \bibnamefont {Timmermans}}, \bibinfo {author} {\bibfnamefont {W.}~\bibnamefont {Ubachs}}, \ and\ \bibinfo {author} {\bibfnamefont {L.}~\bibnamefont {Willmann}},\ }\bibfield  {title} {\enquote {\bibinfo {title} {{Spin-precession method for sensitive electric dipole moment searches}},}\
  }\href@noop {} {\bibfield  {journal} {\bibinfo  {journal} {Physical Review A}\ }\textbf {\bibinfo {volume} {110}},\ \bibinfo {pages} {L010801} (\bibinfo {year} {2024})}\BibitemShut {NoStop}%
\bibitem [{\citenamefont {Hudson}\ \emph {et~al.}(2011)\citenamefont {Hudson}, \citenamefont {Kara}, \citenamefont {Smallman}, \citenamefont {Sauer}, \citenamefont {Tarbutt},\ and\ \citenamefont {Hinds}}]{hudson2011improved}%
  \BibitemOpen
  \bibfield  {author} {\bibinfo {author} {\bibfnamefont {J.~J.}\ \bibnamefont {Hudson}}, \bibinfo {author} {\bibfnamefont {D.~M.}\ \bibnamefont {Kara}}, \bibinfo {author} {\bibfnamefont {I.~J.}\ \bibnamefont {Smallman}}, \bibinfo {author} {\bibfnamefont {B.~E.}\ \bibnamefont {Sauer}}, \bibinfo {author} {\bibfnamefont {M.~R.}\ \bibnamefont {Tarbutt}}, \ and\ \bibinfo {author} {\bibfnamefont {E.~A.}\ \bibnamefont {Hinds}},\ }\bibfield  {title} {\enquote {\bibinfo {title} {Improved measurement of the shape of the electron},}\ }\href@noop {} {\bibfield  {journal} {\bibinfo  {journal} {Nature}\ }\textbf {\bibinfo {volume} {473}},\ \bibinfo {pages} {493} (\bibinfo {year} {2011})}\BibitemShut {NoStop}%
\bibitem [{\citenamefont {Baron}\ \emph {et~al.}(2013)\citenamefont {Baron}, \citenamefont {Campbell}, \citenamefont {DeMille}, \citenamefont {Doyle}, \citenamefont {Gabrielse}, \citenamefont {Gurevich}, \citenamefont {Hess}, \citenamefont {Hutzler}, \citenamefont {Kirilov}, \citenamefont {Kozyryev}, \citenamefont {O{\lq}Leary}, \citenamefont {Panda}, \citenamefont {Parsons}, \citenamefont {Petrik}, \citenamefont {Spaun}, \citenamefont {Vutha},\ and\ \citenamefont {West}}]{baron2013order}%
  \BibitemOpen
  \bibfield  {author} {\bibinfo {author} {\bibfnamefont {J.}~\bibnamefont {Baron}}, \bibinfo {author} {\bibfnamefont {W.~C.}\ \bibnamefont {Campbell}}, \bibinfo {author} {\bibfnamefont {D.}~\bibnamefont {DeMille}}, \bibinfo {author} {\bibfnamefont {J.~M.}\ \bibnamefont {Doyle}}, \bibinfo {author} {\bibfnamefont {G.}~\bibnamefont {Gabrielse}}, \bibinfo {author} {\bibfnamefont {Y.~V.}\ \bibnamefont {Gurevich}}, \bibinfo {author} {\bibfnamefont {P.~W.}\ \bibnamefont {Hess}}, \bibinfo {author} {\bibfnamefont {N.~R.}\ \bibnamefont {Hutzler}}, \bibinfo {author} {\bibfnamefont {E.}~\bibnamefont {Kirilov}}, \bibinfo {author} {\bibfnamefont {I.}~\bibnamefont {Kozyryev}}, \bibinfo {author} {\bibfnamefont {B.~R.}\ \bibnamefont {O{\lq}Leary}}, \bibinfo {author} {\bibfnamefont {C.~D.}\ \bibnamefont {Panda}}, \bibinfo {author} {\bibfnamefont {M.~F.}\ \bibnamefont {Parsons}}, \bibinfo {author} {\bibfnamefont {E.~S.}\ \bibnamefont {Petrik}}, \bibinfo {author} {\bibfnamefont {B.}~\bibnamefont {Spaun}}, \bibinfo {author}
  {\bibfnamefont {A.~C.}\ \bibnamefont {Vutha}}, \ and\ \bibinfo {author} {\bibfnamefont {A.~D.}\ \bibnamefont {West}},\ }\bibfield  {title} {\enquote {\bibinfo {title} {{Order of magnitude smaller limit on the electric dipole moment of the electron}},}\ }\href@noop {} {\bibfield  {journal} {\bibinfo  {journal} {Science}\ ,\ \bibinfo {pages} {1248213}} (\bibinfo {year} {2013})}\BibitemShut {NoStop}%
\bibitem [{\citenamefont {Cairncross}\ \emph {et~al.}(2017)\citenamefont {Cairncross}, \citenamefont {Gresh}, \citenamefont {Grau}, \citenamefont {Cossel}, \citenamefont {Roussy}, \citenamefont {Ni}, \citenamefont {Zhou}, \citenamefont {Ye},\ and\ \citenamefont {Cornell}}]{Cairncross2017}%
  \BibitemOpen
  \bibfield  {author} {\bibinfo {author} {\bibfnamefont {W.~B.}\ \bibnamefont {Cairncross}}, \bibinfo {author} {\bibfnamefont {D.~N.}\ \bibnamefont {Gresh}}, \bibinfo {author} {\bibfnamefont {M.}~\bibnamefont {Grau}}, \bibinfo {author} {\bibfnamefont {K.~C.}\ \bibnamefont {Cossel}}, \bibinfo {author} {\bibfnamefont {T.~S.}\ \bibnamefont {Roussy}}, \bibinfo {author} {\bibfnamefont {Y.}~\bibnamefont {Ni}}, \bibinfo {author} {\bibfnamefont {Y.}~\bibnamefont {Zhou}}, \bibinfo {author} {\bibfnamefont {J.}~\bibnamefont {Ye}}, \ and\ \bibinfo {author} {\bibfnamefont {E.~A.}\ \bibnamefont {Cornell}},\ }\bibfield  {title} {\enquote {\bibinfo {title} {{Precision measurement of the electron's electric dipole moment using trapped molecular ions}},}\ }\href@noop {} {\bibfield  {journal} {\bibinfo  {journal} {Physical Review Letters}\ }\textbf {\bibinfo {volume} {119}},\ \bibinfo {pages} {153001} (\bibinfo {year} {2017})}\BibitemShut {NoStop}%
\bibitem [{\citenamefont {Andreev}\ \emph {et~al.}(2018)\citenamefont {Andreev}, \citenamefont {Ang}, \citenamefont {DeMille}, \citenamefont {Gabrielse}, \citenamefont {Haefner}, \citenamefont {Hutzler}, \citenamefont {Lasner}, \citenamefont {Meisenhelder}, \citenamefont {O’Leary}, \citenamefont {Panda}, \citenamefont {West}, \citenamefont {West},\ and\ \citenamefont {Wu}}]{acme2018improved}%
  \BibitemOpen
  \bibfield  {author} {\bibinfo {author} {\bibfnamefont {V.}~\bibnamefont {Andreev}}, \bibinfo {author} {\bibfnamefont {D.~G.}\ \bibnamefont {Ang}}, \bibinfo {author} {\bibfnamefont {J.~M.}\ \bibnamefont {DeMille}, \bibfnamefont {D.~Doyle}}, \bibinfo {author} {\bibfnamefont {G.}~\bibnamefont {Gabrielse}}, \bibinfo {author} {\bibfnamefont {J.}~\bibnamefont {Haefner}}, \bibinfo {author} {\bibfnamefont {N.~R.}\ \bibnamefont {Hutzler}}, \bibinfo {author} {\bibfnamefont {Z.}~\bibnamefont {Lasner}}, \bibinfo {author} {\bibfnamefont {C.}~\bibnamefont {Meisenhelder}}, \bibinfo {author} {\bibfnamefont {B.~R.}\ \bibnamefont {O’Leary}}, \bibinfo {author} {\bibfnamefont {C.~D.}\ \bibnamefont {Panda}}, \bibinfo {author} {\bibfnamefont {A.~D.}\ \bibnamefont {West}}, \bibinfo {author} {\bibfnamefont {E.~P.}\ \bibnamefont {West}}, \ and\ \bibinfo {author} {\bibfnamefont {X.}~\bibnamefont {Wu}},\ }\bibfield  {title} {\enquote {\bibinfo {title} {{ Improved limit on the electric dipole moment of the electron}},}\ }\href@noop
  {} {\bibfield  {journal} {\bibinfo  {journal} {Nature}\ }\textbf {\bibinfo {volume} {562}},\ \bibinfo {pages} {355--360} (\bibinfo {year} {2018})}\BibitemShut {NoStop}%
\bibitem [{\citenamefont {Roussy}\ \emph {et~al.}(2023)\citenamefont {Roussy}, \citenamefont {Caldwell}, \citenamefont {Wright}, \citenamefont {Cairncross}, \citenamefont {Shagam}, \citenamefont {Ng}, \citenamefont {Schlossberger}, \citenamefont {Park}, \citenamefont {Wang}, \citenamefont {Ye},\ and\ \citenamefont {Cornell}}]{roussy2023new}%
  \BibitemOpen
  \bibfield  {author} {\bibinfo {author} {\bibfnamefont {T.~S.}\ \bibnamefont {Roussy}}, \bibinfo {author} {\bibfnamefont {L.}~\bibnamefont {Caldwell}}, \bibinfo {author} {\bibfnamefont {T.}~\bibnamefont {Wright}}, \bibinfo {author} {\bibfnamefont {W.~B.}\ \bibnamefont {Cairncross}}, \bibinfo {author} {\bibfnamefont {Y.}~\bibnamefont {Shagam}}, \bibinfo {author} {\bibfnamefont {K.~B.}\ \bibnamefont {Ng}}, \bibinfo {author} {\bibfnamefont {N.}~\bibnamefont {Schlossberger}}, \bibinfo {author} {\bibfnamefont {S.~Y.}\ \bibnamefont {Park}}, \bibinfo {author} {\bibfnamefont {A.}~\bibnamefont {Wang}}, \bibinfo {author} {\bibfnamefont {J.}~\bibnamefont {Ye}}, \ and\ \bibinfo {author} {\bibfnamefont {E.~A.}\ \bibnamefont {Cornell}},\ }\bibfield  {title} {\enquote {\bibinfo {title} {A new bound on the electron's electric dipole moment},}\ }\href@noop {} {\bibfield  {journal} {\bibinfo  {journal} {Science}\ }\textbf {\bibinfo {volume} {381}},\ \bibinfo {pages} {46} (\bibinfo {year} {2023})}\BibitemShut {NoStop}%
\bibitem [{\citenamefont {Lambo}\ \emph {et~al.}(2023{\natexlab{a}})\citenamefont {Lambo}, \citenamefont {Koyanagi}, \citenamefont {Ragyanszki}, \citenamefont {Horbatsch}, \citenamefont {Fournier},\ and\ \citenamefont {Hessels}}]{lambo2023calculationAr}%
  \BibitemOpen
  \bibfield  {author} {\bibinfo {author} {\bibfnamefont {R.~L.}\ \bibnamefont {Lambo}}, \bibinfo {author} {\bibfnamefont {G.~K.}\ \bibnamefont {Koyanagi}}, \bibinfo {author} {\bibfnamefont {A.}~\bibnamefont {Ragyanszki}}, \bibinfo {author} {\bibfnamefont {M.}~\bibnamefont {Horbatsch}}, \bibinfo {author} {\bibfnamefont {R.}~\bibnamefont {Fournier}}, \ and\ \bibinfo {author} {\bibfnamefont {E.~A.}\ \bibnamefont {Hessels}},\ }\bibfield  {title} {\enquote {\bibinfo {title} {{Calculation of the local environment of a barium monofluoride molecule in an argon matrix: a step towards using matrix-isolated BaF for determining the electron electric dipole moment}},}\ }\href@noop {} {\bibfield  {journal} {\bibinfo  {journal} {Molecular Physics}\ }\textbf {\bibinfo {volume} {121}},\ \bibinfo {pages} {e2198044} (\bibinfo {year} {2023}{\natexlab{a}})}\BibitemShut {NoStop}%
\bibitem [{\citenamefont {Lambo}\ \emph {et~al.}(2023{\natexlab{b}})\citenamefont {Lambo}, \citenamefont {Koyanagi}, \citenamefont {Horbatsch}, \citenamefont {Fournier},\ and\ \citenamefont {Hessels}}]{lambo2023calculationNe}%
  \BibitemOpen
  \bibfield  {author} {\bibinfo {author} {\bibfnamefont {R.~L.}\ \bibnamefont {Lambo}}, \bibinfo {author} {\bibfnamefont {G.~K.}\ \bibnamefont {Koyanagi}}, \bibinfo {author} {\bibfnamefont {M.}~\bibnamefont {Horbatsch}}, \bibinfo {author} {\bibfnamefont {R.}~\bibnamefont {Fournier}}, \ and\ \bibinfo {author} {\bibfnamefont {E.~A.}\ \bibnamefont {Hessels}},\ }\bibfield  {title} {\enquote {\bibinfo {title} {{Calculation of the local environment of a barium monofluoride molecule in a neon matrix}},}\ }\href@noop {} {\bibfield  {journal} {\bibinfo  {journal} {Molecular Physics}\ }\textbf {\bibinfo {volume} {121}},\ \bibinfo {pages} {e2232051} (\bibinfo {year} {2023}{\natexlab{b}})}\BibitemShut {NoStop}%
\bibitem [{\citenamefont {Li}\ \emph {et~al.}(2023)\citenamefont {Li}, \citenamefont {Ramachandran}, \citenamefont {Anderson},\ and\ \citenamefont {Vutha}}]{li2023optical}%
  \BibitemOpen
  \bibfield  {author} {\bibinfo {author} {\bibfnamefont {S.~J.}\ \bibnamefont {Li}}, \bibinfo {author} {\bibfnamefont {H.~D.}\ \bibnamefont {Ramachandran}}, \bibinfo {author} {\bibfnamefont {R.}~\bibnamefont {Anderson}}, \ and\ \bibinfo {author} {\bibfnamefont {A.~C.}\ \bibnamefont {Vutha}},\ }\bibfield  {title} {\enquote {\bibinfo {title} {{Optical control of BaF molecules trapped in neon ice}},}\ }\href@noop {} {\bibfield  {journal} {\bibinfo  {journal} {New Journal of Physics}\ }\textbf {\bibinfo {volume} {25}},\ \bibinfo {pages} {082001} (\bibinfo {year} {2023})}\BibitemShut {NoStop}%
\bibitem [{\citenamefont {Koyanagi}\ \emph {et~al.}(2023)\citenamefont {Koyanagi}, \citenamefont {Lambo}, \citenamefont {Ragyanszki}, \citenamefont {Fournier}, \citenamefont {Horbatsch},\ and\ \citenamefont {Hessels}}]{koyanagi2023accurate}%
  \BibitemOpen
  \bibfield  {author} {\bibinfo {author} {\bibfnamefont {G.~K.}\ \bibnamefont {Koyanagi}}, \bibinfo {author} {\bibfnamefont {R.~L.}\ \bibnamefont {Lambo}}, \bibinfo {author} {\bibfnamefont {A.}~\bibnamefont {Ragyanszki}}, \bibinfo {author} {\bibfnamefont {R.}~\bibnamefont {Fournier}}, \bibinfo {author} {\bibfnamefont {M.}~\bibnamefont {Horbatsch}}, \ and\ \bibinfo {author} {\bibfnamefont {E.~A.}\ \bibnamefont {Hessels}},\ }\bibfield  {title} {\enquote {\bibinfo {title} {{Accurate calculation of the interaction of a barium monofluoride molecule with an argon atom: A step towards using matrix isolation of BaF for determining the electron electric dipole moment}},}\ }\href@noop {} {\bibfield  {journal} {\bibinfo  {journal} {Journal of Molecular Spectroscopy}\ }\textbf {\bibinfo {volume} {391}},\ \bibinfo {pages} {111736} (\bibinfo {year} {2023})}\BibitemShut {NoStop}%
\bibitem [{\citenamefont {Truppe}\ \emph {et~al.}(2018)\citenamefont {Truppe}, \citenamefont {Hambach}, \citenamefont {Skoff}, \citenamefont {Bulleid}, \citenamefont {Bumby}, \citenamefont {Hendricks}, \citenamefont {Hinds}, \citenamefont {Sauer},\ and\ \citenamefont {Tarbutt}}]{truppe2018buffer}%
  \BibitemOpen
  \bibfield  {author} {\bibinfo {author} {\bibfnamefont {S.}~\bibnamefont {Truppe}}, \bibinfo {author} {\bibfnamefont {M.}~\bibnamefont {Hambach}}, \bibinfo {author} {\bibfnamefont {S.~M.}\ \bibnamefont {Skoff}}, \bibinfo {author} {\bibfnamefont {N.~E.}\ \bibnamefont {Bulleid}}, \bibinfo {author} {\bibfnamefont {J.~S.}\ \bibnamefont {Bumby}}, \bibinfo {author} {\bibfnamefont {R.~J.}\ \bibnamefont {Hendricks}}, \bibinfo {author} {\bibfnamefont {E.~A.}\ \bibnamefont {Hinds}}, \bibinfo {author} {\bibfnamefont {B.~E.}\ \bibnamefont {Sauer}}, \ and\ \bibinfo {author} {\bibfnamefont {M.~R.}\ \bibnamefont {Tarbutt}},\ }\bibfield  {title} {\enquote {\bibinfo {title} {{A buffer gas beam source for short, intense and slow molecular pulses}},}\ }\href@noop {} {\bibfield  {journal} {\bibinfo  {journal} {Journal of Modern Optics}\ }\textbf {\bibinfo {volume} {65}},\ \bibinfo {pages} {648} (\bibinfo {year} {2018})}\BibitemShut {NoStop}%
\bibitem [{\citenamefont {Yau}\ \emph {et~al.}(2024)\citenamefont {Yau}, \citenamefont {Corriveau}, \citenamefont {McCall}, \citenamefont {Perez~Garcia}, \citenamefont {Heinrich}, \citenamefont {Lambo}, \citenamefont {Koyanagi}, \citenamefont {Chauhan}, \citenamefont {George}, \citenamefont {Storry}, \citenamefont {Horbatsch},\ and\ \citenamefont {Hessels}}]{yau2024specular}%
  \BibitemOpen
  \bibfield  {author} {\bibinfo {author} {\bibfnamefont {H.-M.}\ \bibnamefont {Yau}}, \bibinfo {author} {\bibfnamefont {Z.}~\bibnamefont {Corriveau}}, \bibinfo {author} {\bibfnamefont {N.}~\bibnamefont {McCall}}, \bibinfo {author} {\bibfnamefont {J.}~\bibnamefont {Perez~Garcia}}, \bibinfo {author} {\bibfnamefont {D.}~\bibnamefont {Heinrich}}, \bibinfo {author} {\bibfnamefont {R.~L.}\ \bibnamefont {Lambo}}, \bibinfo {author} {\bibfnamefont {G.~K.}\ \bibnamefont {Koyanagi}}, \bibinfo {author} {\bibfnamefont {T.}~\bibnamefont {Chauhan}}, \bibinfo {author} {\bibfnamefont {M.~C.}\ \bibnamefont {George}}, \bibinfo {author} {\bibfnamefont {C.~H.}\ \bibnamefont {Storry}}, \bibinfo {author} {\bibfnamefont {M.}~\bibnamefont {Horbatsch}}, \ and\ \bibinfo {author} {\bibfnamefont {E.~A.}\ \bibnamefont {Hessels}},\ }\bibfield  {title} {\enquote {\bibinfo {title} {{Specular reflection of polar molecules from a simple multi-cylinder electrostatic mirror: a method for separating BaF molecules produced in a buffer-gas-cooled
  laser-ablation source from other ablation products}},}\ }\href@noop {} {\bibfield  {journal} {\bibinfo  {journal} {Arxiv}\ } (\bibinfo {year} {2024})}\BibitemShut {NoStop}%
\bibitem [{\citenamefont {Ramsthaler-Sommer}\ \emph {et~al.}(1986)\citenamefont {Ramsthaler-Sommer}, \citenamefont {Eberhardt},\ and\ \citenamefont {Schurath}}]{ramsthaler1986radiative}%
  \BibitemOpen
  \bibfield  {author} {\bibinfo {author} {\bibfnamefont {A.}~\bibnamefont {Ramsthaler-Sommer}}, \bibinfo {author} {\bibfnamefont {K.~E.}\ \bibnamefont {Eberhardt}}, \ and\ \bibinfo {author} {\bibfnamefont {U.}~\bibnamefont {Schurath}},\ }\bibfield  {title} {\enquote {\bibinfo {title} {{Radiative decay and radiationless relaxation of NH/ND ($a\,^1\Delta$) isolated in rare gas matrices}},}\ }\href@noop {} {\bibfield  {journal} {\bibinfo  {journal} {The Journal of Chemical Physics}\ }\textbf {\bibinfo {volume} {85}},\ \bibinfo {pages} {3760} (\bibinfo {year} {1986})}\BibitemShut {NoStop}%
\end{thebibliography}%

% \newpage

% \section*{Supplementary Materials}

% \begin{figure}[b!]
% \includegraphics[width=3.4in]
% {HeFSCompareFigcropped.pdf}
% \caption{\label{fig:history}
% The history of improvement 
% % \textcolor{red}{\sout{of}}
% in precision of  
% measurements and theory for the 
% helium
% 2$^3$P 
% fine structure.
% The 
% exes 
% % \textcolor{red}{
% (shown in various colors
% for clarity of labeling)
% % }
% indicate the quoted
% uncertainty 
% in parts per billion (ppb)
% for measurements
% of the 
% $J$$=$0-to-$J$$=$1
% and
% $J$$=$0-to-$J$$=$2
% intervals
% performed since
% 1965. 
% For measurements that
% disagree with 
% % \textcolor{red}{
% % \sout{the current result}
% our new result,
% % }, 
% the number of standard
% deviations for this
% disagreement is indicated,
% and the arrows show
% the expanded uncertainty
% that would be required 
% for agreement.
% The purple diamonds show 
% the 
% difference between
% theory and experiment,
% which has improved 
% with the 
% completion of 
% calculations of all
% contributions 
% of order 
% $\alpha^6$ 
% \cite{PRL.29.12}
% and 
% $\alpha^7$
% \cite{PRL.104.070403}.
% Calculations of 
% $\alpha^8$
% terms will allow
% for a 
% ppb-level
% comparison between
% experiment and theory.
% }
% \end{figure}

% \begin{figure}[b!]
% \includegraphics[width=2.3 in]
% {alphaComparison.pdf}
% \caption{\label{fig:alpha}
% Current 
% determinations of $\alpha$
% obtained from atomic recoil 
% in 
% Cs
% and 
% Rb
% and from 
% the electron 
% magnetic dipole moment
% ($g_e$), 
% which span a range
% of over 
% 1~ppb.
% }
% \end{figure}

\end{document}